\documentclass[conference, 9pt]{IEEEtran}

\usepackage{times}
\usepackage{subfigure}
\usepackage{graphicx}
\usepackage{amsmath}
\usepackage{bm}
\usepackage{url}
\usepackage{stmaryrd}
\usepackage{balance}
\usepackage{amssymb}
\usepackage{pgfplots}
\usepackage{pifont}
\usepackage{epsfig}
\usepackage{booktabs}
\usepackage{pgffor}
\usepackage{tikz}
\usepackage{multirow}
\usepackage{amsmath}

\makeatletter
\newcommand\xleftrightarrow[2][]{%
  \ext@arrow 9999{\longleftrightarrowfill@}{#1}{#2}}
\newcommand\longleftrightarrowfill@{%
  \arrowfill@\leftarrow\relbar\rightarrow}
\makeatother

\newcommand{\mb}{\mathbf}

\newcommand{\problem}{\textsc{BL-MNE}}

\newcommand{\our}{\textsc{DIME}}
\newcommand{\oursingle}{\textsc{DIME-SH}}
\newcommand{\autoencoder}{\textsc{Auto-Encoder}}
\newcommand{\linemodel}{\textsc{LINE}}
\newcommand{\deepwalk}{\textsc{DeepWalk}}

\usepackage{algorithm}
\usepackage{algorithmic}

\begin{document}
\title{BL-MNE: Emerging Heterogeneous Social Network Embedding through Broad Learning with Aligned Autoencoder}


\author{
\IEEEauthorblockN{Jiawei Zhang$^\star$, Congying Xia$^\S$, Chenwei Zhang$^\S$, Limeng Cui$^\dagger$, Yanjie Fu$^\ddagger$ and Philip S. Yu$^{\S,\P}$}
\IEEEauthorblockA{$^\star$IFM Lab, Department of Computer Science, Florida State University, FL, USA\\
			      $^\S$University of Illinois at Chicago, Chicago, IL, USA\\
			      $^\dagger$School of Computer and Control Engineering, University of Chinese Academy of Sciences, Beijing, China\\
			      $^\ddagger$Missouri University of Science and Technology, MO, USA\\
			      $^\P$Shanghai Institute for Advanced Communication and Data Science, Fudan University, Shanghai, China\\
			      {jzhang@cs.fsu.edu, \{cxia8, czhang99, psyu\}@uic.edu, lmcui932@163.com, fuyan@mst.edu}
			      }
}

\maketitle

\begin{abstract}

Network embedding aims at projecting the network data into a low-dimensional feature space, where the nodes are represented as a unique feature vector and network structure can be effectively preserved. In recent years, more and more online application service sites can be represented as massive and complex networks, which are extremely challenging for traditional machine learning algorithms to deal with. Effective embedding of the complex network data into low-dimension feature representation can both save data storage space and enable traditional machine learning algorithms applicable to handle the network data. Network embedding performance will degrade greatly if the networks are of a sparse structure, like the emerging networks with few connections. In this paper, we propose to learn the embedding representation for a target emerging network based on the broad learning setting, where the emerging network is aligned with other external mature networks at the same time. To solve the problem, a new embedding framework, namely ``\underline{D}eep al\underline{I}gned autoencoder based e\underline{M}b\underline{E}dding'' (\our), is introduced in this paper. {\our} handles the diverse link and attribute in a unified analytic based on broad learning, and introduces the multiple aligned attributed heterogeneous social network concept to model the network structure. A set of meta paths are introduced in the paper, which define various kinds of connections among users via the heterogeneous link and attribute information. The closeness among users in the networks are defined as the meta proximity scores, which will be fed into {\our} to learn the embedding vectors of users in the emerging network. Extensive experiments have been done on real-world aligned social networks, which have demonstrated the effectiveness of {\our} in learning the emerging network embedding vectors.

\end{abstract}



\section{Introduction}\label{sec:introduction}

In the era of big data, a rapidly increasing number of online application websites appear recently, which can be represented as massive and complex networks. The representative examples include \textit{online social networks}, like Facebook and Twitter, \textit{e-commerce sites}, like Amazon and eBay, \textit{academic sites}, like DBLP and Google Scholar, as well as \textit{POIs recommendation sites}, like Foursquare and Yelp. These network data can be very difficult to deal with due to their \textit{extremely large scale} (involving millions even billions of nodes), \textit{complex structures} (containing heterogeneous links) as well as \textit{diverse attributes} (attached to the nodes or links). Great challenges exist in handling these complex network representation data with traditional machine learning algorithms, which usually take feature vectors as the input and cannot handle graph data directly. A general representation of heterogeneous networks as feature vectors is desired for knowledge discovery from such complex-structured data. In this paper, we will use online social networks as the example to illustrate the studied problem as well as the learning framework.

In recent years, many research works propose to embed the online social network data into a low-dimensional feature space \cite{PAS14, TQWZYM15, GL16}, in which each node is represented as a unique feature vector. From these feature vectors, the original network structure can be effectively reconstructed. With these embedded feature vectors, classic machine learning algorithms can be applied to deal with the social network data directly, and the storage space can also be saved greatly. However, most existing social network embedding methods are proposed for homogeneous networks, which learn the feature vectors for user nodes merely based on the social connections among them. When applied to handle real-world social network data, these embedding models can hardly work well. The main reason is that the internal social links are usually very sparse in online social networks \cite{TQWZYM15}, which can hardly preserve the complete network structure. For a pair of users who are not directed connected, these models will not be able determine the closeness of these users' feature vectors in the embedding space. Such a problem will be more severe when it comes to the \textit{emerging social networks} \cite{ZY15}, which denote the newly created online social networks with very few social connections.

\begin{table*}[t]
\vspace{-30pt}
\centering
\caption{Summary of related problems.}\label{tab:related}
\begin{tabular}{|l|l|l|l|l|}
\hline
&\textbf{Aligned Heterogeneous}
&\textbf{Translation based Graph}
&\textbf{Homogeneous Network}
&\textbf{Heterogeneous Network}\\
Property	
&\textbf{Network Embedding}
&\textbf{Embedding} \cite{BUGWY13, WZFC14, LLSLZ15}
&\textbf{Embedding} \cite{PAS14, TQWZYM15, GL16}
&\textbf{Embedding} \cite{CHTQAH15, CS16}  \\
\hline
target network	&emerging		&regular			&regular			&regular\\
network		&attributed heterogeneous	&multi-relational	&homogeneous		&heterogeneous\\
\#network		&multiple			&single			&single			&single\\
proximity		&meta proximity	&first order			&first/second order/random walk	&first order \cite{CHTQAH15}, meta path \cite{CS16}\\
multi-source fusion	&anchor link based fusion		&N/A				&N/A				&N/A\\
\hline
\end{tabular}\vspace{-12pt}
\end{table*}

Meanwhile, as discovered in \cite{ZYZ14}, to enjoy more social network services, people nowadays are usually involved in multiple online social networks at the same time. For instance, people tend to join in Facebook for casual socialization with their classmates; they will use Foursquare to search for nearby restaurants for dinner; and they will turn to use Instagram to share photos with their friends online. Users who are involved in these emerging social networks may have been using other well-developed social networks (e.g., Facebook, Twitter) for a long time. Information available for the users in other aligned mature networks is usually very abundant and of a sufficient amount. Effective information exchanges from these mature networks to the emerging networks for the shared users can help overcome the information sparsity problem promisingly, which is a important topic covered in the broad learning task \cite{ZCYLF17, ZZHWZZY17} to be introduced as follows. To denote the accounts owned by the same people in different online social networks, an \textit{anchor link} will be added to connect their account pair between the networks \cite{ZYZ14}. Formally, the online social networks connected by the anchor links (between the shared user accounts) are called \textit{multiple aligned social networks} \cite{ZYZ14}.

\noindent \textbf{Problem Studied}: In this paper, we propose to study the emerging network embedding problem across multiple aligned heterogeneous social networks simultaneously based on the broad learning setting, which is formally named as the ``\underline{B}road \underline{L}earning based e\underline{M}erging \underline{N}etwork \underline{E}mbedding'' ({\problem}) problem. In the concurrent embedding process based on the broad learning setting, {\problem} aims at distilling relevant information from both the emerging and other aligned mature networks to derive compliment knowledge and learn a good vector representation for user nodes in the emerging network.

Here, ``\textbf{Broad Learning}'' \cite{ZCYLF17, ZZHWZZY17} is a new type of learning task, which focuses on fusing multiple large-scale information sources of diverse varieties together and carrying out synergistic data mining tasks across these fused sources in one unified analytic \cite{ZY15-4, ZCZCYH17, ZYZ14, ZY15, ZYL15, ZLY15, ZYLZ16, ZYL17, ZYL17-2}. In the real world, on the same information entities, e.g., social media users \cite{ZY15-4, ZCZCYH17, ZYZ14, ZY15}, movie knowledge library entries \cite{ZZHWZZY17} and employees in companies \cite{ZYL15, ZLY15, ZYLZ16, ZYL17, ZYL17-2}, a large amount of information can actually be collected from various sources. These sources are usually of different varieties, like Foursquare vs Twitter \cite{ZY15-4, ZCZCYH17, ZYZ14, ZY15}, IMDB vs Douban Movie sites \cite{ZZHWZZY17}, Yammer vs company organizational chart \cite{ZYL15, ZLY15, ZYLZ16, ZYL17, ZYL17-2}. Each information source provides a specific signature of the same entity from a unique underlying aspect. Effective fusion of these different information sources provides an opportunity for researchers and practitioners to understand the entities more comprehensively, which renders ``\textbf{Broad Learning}'' an extremely important learning task. Fusing and mining multiple information sources of large volumes and diverse varieties are the fundamental problems in big data studies. ``\textbf{Broad Learning}''  investigates the principles, methodologies and algorithms for synergistic knowledge discovery across multiple information sources, and evaluates the corresponding benefits \cite{ZCYLF17, ZZHWZZY17}. Great challenges exist in ``\textbf{Broad Learning}'' for the effective fusion of relevant knowledge across different aligned information sources depends upon not only the relatedness of these information sources, but also the target application problems. ``\textbf{Broad Learning}'' aims at developing general methodologies, which will be shown to work for a diverse set of applications, while the specific parameter settings can be learned for each application from the training data \cite{ZCYLF17, ZZHWZZY17}.

{\problem} is significantly different from existing network embedding problems \cite{BUGWY13, WZFC14, PAS14, LLSLZ15, TQWZYM15, CHTQAH15, GL16, CS16} in several perspectives. First of all, the target network studied in {\problem} is an emerging network suffering from the information sparsity problem, which is different from the embedding problems for regular networks \cite{BUGWY13, WZFC14, LLSLZ15, CHTQAH15, CS16}. Secondly, the networks studied in {\problem} are all heterogeneous networks containing complex links and diverse attributes, which renders {\problem} different from existing homogeneous network embedding problems \cite{PAS14, TQWZYM15, GL16}. Furthermore, {\problem} is based on the multiple aligned networks setting, where information from aligned networks will be exchanged to refine the embedding results mutually, and it is different from the existing single-network based embedding problems \cite{BUGWY13, WZFC14, PAS14, LLSLZ15, CHTQAH15, TQWZYM15, CHTQAH15, GL16, CS16}. We also provide a summary about the difference between {\problem} and existing works in Table~\ref{tab:related} (which summarizes and compares several related works in different aspects), and more information about other related works will be introduced Section~\ref{sec:related_work} at the end of the paper.

The {\problem} problem is not an easy problem, and it has several great challenges to deal with, which are provided as follows:

\begin{itemize}

\item \textit{Problem Formulation}: To overcome the information sparsity problem, {\problem} studies the concurrent embedding of multiple aligned social networks, which is still an open problem to this context so far. Formal definition and formulation of the {\problem} problem is required before we introduce the solutions.

\item \textit{Heterogeneity of Networks}: The networks studied in this paper are of very complex structures. Besides the regular social connections among users, there also exist many other types of links as well as diverse attributes attached to the user nodes. Effective incorporating these heterogeneous information into a unified embedding analytic is a great challenge.

\item \textit{Multiple Aligned Network Embedding Framework}: Due to the significant differences between {\problem} with the existing works, few existing network embedding models can be applied to address the {\problem} directly. A new embedding learning framework is needed to learn the emerging network embedding vectors across multiple aligned networks synergistically.

\end{itemize}

To address all these challenges aforementioned, in this paper, we introduce a novel multiple aligned heterogeneous social network embedding framework, named ``\underline{D}eep al\underline{I}gned autoencoder based e\underline{M}b\underline{E}dding'' (\our). To handle the heterogeneous link and attribute information in a unified analytic, we introduce the \textit{aligned attribute augmented heterogeneous network} concept in this paper. From these heterogeneous networks a set of meta paths are introduced to represent the diverse connections among users in online social networks (via social links, other diverse connections, and various attributes). A set of \textit{meta proximity} measures are defined for each of the meta paths denoting the closeness among users. The meta proximity information will be fed into a deep learning framework, which takes the input information from multiple aligned heterogeneous social networks simultaneously, to achieve the embedding feature vectors for all the users in these aligned networks. Based on the connection among users, framework {\our} aims at embedding close user nodes to a close area in the low-dimensional feature space for each of the social networks respectively. Meanwhile, framework {\our} also poses constraints on the feature vectors corresponding to the shared users across networks to map them to a relatively close region. In this way, information can be transferred from the mature networks to the emerging network and solve the \textit{information sparsity} problem.

The remaining parts of this paper are organized as follows. We will provide the terminology definition and problem formulation in Section~\ref{sec:formulation}. Information about the framework is available in Section~\ref{sec:method}, which will be evaluated in Section~\ref{sec:experiment}. Finally, we will introduce the related works in Section~\ref{sec:related_work} and conclude this paper in Section~\ref{sec:conclusion}.

\section{Terminology Definition and Problem Formulation} \label{sec:formulation}

In this section, we will first introduce the definitions of several important terminologies, based on which we will then provide the formulation of the {\problem} problem.

\subsection{Terminology Definition}

The social networks studied in this paper contain different categories of nodes and links, as well as very diverse attributes attached to the nodes. Formally, we can represent these network structured data as the \textit{attributed heterogeneous social networks}.

\noindent \textbf{Definition 1} (Attributed Heterogeneous Social Networks): The \textit{attributed heterogeneous social network} can be represented as a graph $G = (\mathcal{V}, \mathcal{E}, \mathcal{T})$, where $\mathcal{V} = \bigcup_i \mathcal{V}_i$ denotes the set of nodes belonging to various categories and $\mathcal{E} = \bigcup_i \mathcal{E}_i$ represents the set of diverse links among the nodes. What's more, $\mathcal{T}$ = $\bigcup_i \mathcal{T}_i$ denotes the set of attributes attached to the nodes in $\mathcal{V}$. For user $u$ in the network, we can represent the $i_{th}$ type of attribute associated to $u$ as $T_i(u)$, and all the attributes $u$ has can be represented as $T(u) = \bigcup_i {T}_i(u)$.

The social network datasets used in this paper include Foursquare and Twitter. Formally, the Foursquare and Twitter can both be represented as the \textit{attributed heterogeneous social networks} $G = (\mathcal{V}, \mathcal{E}, \mathcal{T})$, where $\mathcal{V} = \mathcal{U} \cup \mathcal{P}$ involves the user and post nodes, and $\mathcal{E} = \mathcal{E}_{u,u} \cup \mathcal{E}_{u,p}$ contains the links among users and those between users and posts. In addition, the nodes in $\mathcal{V}$ are also attached with a set of attributes, i.e., $\mathcal{T}$. For instance, for the posts written by users, we can obtain the contained textual contents, timestamps and checkins, which can all be represented as the attributes of the post nodes.

Between Foursquare and Twitter, there may exist a large number of shared common users, who can align the networks together. In this paper, we will follow the concept definitions proposed in \cite{ZYZ14}, and call the user account correspondence relationships as the \textit{anchor links}. Meanwhile, the networks connected by the \textit{anchor links} are called the multiple \textit{aligned attributed heterogeneous social networks} (or \textit{aligned social networks} for short).

\noindent \textbf{Definition 2} (Multiple Aligned Social Networks): Formally, given $n$ attributed heterogeneous social networks $\{G^{(1)}, \cdots, G^{(n)}\}$ with shared users, they can be defined as \textit{multiple aligned social networks} $\mathcal{G} = ((G^{(1)}, \cdots, G^{(n)}), (\mathcal{A}^{(1,2)}, \cdots, \mathcal{A}^{(n-1, n)}))$. Set $\mathcal{A}^{(i,j)}$ represents the anchor links between $G^{(i)}$ and $G^{(j)}$. User pair $(u^{(i)}, v^{(j)}) \in \mathcal{A}^{(i,j)}$ iff $u^{(i)}$ and $v^{(j)}$ are the accounts of the same user in networks $G^{(i)}$ and $G^{(j)}$ respectively.

For the Foursquare and Twitter social networks used in this paper, we can represent them as two aligned social networks $\mathcal{G} = ((G^{(1)}, G^{(2)}), (\mathcal{A}^{(1,2)}))$, which will be used as an example to illustrate the models. A simple extension of the proposed framework can be applied to \textit{k aligned networks} very easily.

\subsection{Problem Formulation}

\noindent \textbf{Problem Definition} ({\problem} Problem): Given two aligned networks $\mathcal{G} = ((G^{(1)}, G^{(2)}), (\mathcal{A}^{(1,2)}))$, where $G^{(1)}$ is an emerging network and $G^{(2)}$ is a mature network, {\problem} aims at learning a mapping function $f^{(i)}: \mathcal{U}^{(i)} \to \mathbb{R}^{d^{(i)}}$ to project the user node in $G^{(i)}$ to a feature space of dimension $d^{(i)}$ ($d^{(i)} \ll |\mathcal{U}|^{(i)}$). The objective of mapping functions $f^{(i)}$ is to ensure the embedding results can preserve the network structural information, where similar user nodes will be projected to close regions. Furthermore, in the embedding process, {\problem} also wants to transfer information between $G^{(2)}$ and $G^{(1)}$ to overcome the information sparsity problem in $G^{(1)}$.


\section{Proposed Method}\label{sec:method}

In this section, we will introduce the framework {\our} in detail. At the beginning, we provide the notations used in the paper. After that, in Section~\ref{subsec:heterogeneous_proximity}, we will talk about how to calculate the \textit{meta proximity} scores among users based on information in the attributed heterogeneous social networks. With the \textit{meta proximity} measures, the {\our} framework will be introduced in Section~\ref{subsec:deep} to obtain the embedding vectors of user nodes across aligned networks, where information from other aligned mature networks will be used to refine the embedding vectors in the emerging sparse network.

\subsection{Notations}\label{subsec:notation}

In the sequel, we will use the lower case letters (e.g., $x$) to represent scalars, lower case bold letters (e.g., $\mb{x}$) to denote column vectors, bold-face upper case letters (e.g., $\mb{X}$) to denote matrices, and upper case calligraphic letters (e.g., $\mathcal{X}$) to denote sets. Given a matrix $\mb{X}$, we denote $\mb{X}(i,:)$ and $\mb{X}(:,j)$ as the $i_{th}$ row and $j_{th}$ column of matrix $\mb{X}$ respectively. The ($i_{th}$, $j_{th}$) entry of matrix $\mb{X}$ can be denoted as either $X(i,j)$ or $X_{i,j}$, which will be used interchangeably in this paper. We use $\mb{X}^\top$ and $\mb{x}^\top$ to represent the transpose of matrix $\mb{X}$ and vector $\mb{x}$. For vector $\mb{x}$, we represent its $L_p$-norm as $\left\| \mb{x} \right\|_p = (\sum_i |x_i|^p)^{\frac{1}{p}}$. The $L_p$-norm of matrix $\mb{X}$ can be represented as $\left\| \mb{X} \right\|_p = (\sum_{i,j} |X_{i,j}|^p)^{\frac{1}{p}}$. The element-wise product of vectors $\mb{x}$ and $\mb{y}$ of the same dimension is represented as $\mb{x} \odot \mb{y}$, while the element-wise product of matrices $\mb{X}$ and $\mb{Y}$ of the same dimensions is represented as $\mb{X} \odot \mb{Y}$.

\begin{table*}[t]
\vspace{-30pt}
\scriptsize
\centering
{
\caption{Summary of Social Meta Paths (for both Foursquare and Twitter).}\label{tab:meta_path}
\begin{tabular}{llll}
\hline
\textbf{ID}
&\textbf{Notation}
& \textbf{Heterogeneous Network Meta Path}
& \textbf{Semantics}\\
\hline
\hline

$\Phi_0$
&U $\to$ U
&User $\xrightarrow{follow}$ User
&Follow\\

$\Phi_1$
&U $\to$ U $\to$ U
&User $\xrightarrow{follow}$ User $\xrightarrow{follow}$ User
&Follower of Follower\\

$\Phi_2$
&U $\to$ U $\gets$ U
&User $\xrightarrow{follow}$ User $\xrightarrow{follow^{-1}}$ User
&Common Out Neighbor\\

$\Phi_3$
&U $\gets$ U $\to$ U
&User $\xrightarrow{follow^{-1}}$ User $\xrightarrow{follow}$ User
&Common In Neighbor\\

$\Phi_4$
&U $\gets$ U $\gets$ U
&User $\xrightarrow{follow^{-1}}$ User $\xrightarrow{follow^{-1}}$ User
&Followee of Followee\\

\hline


$\Phi_5$
&U $\to$ P $\to$ W $\gets$ P $\gets$ U
&User $\xrightarrow{write}$ Post $\xrightarrow{have}$ Word $\xrightarrow{have^{-1}}$ Post $\xrightarrow{write^{-1}}$ User
&Posts Containing Common Words\\


$\Phi_6$
&U $\to$ P $\to$ T $\gets$ P $\gets$ U
&User $\xrightarrow{write}$ Post $\xrightarrow{have}$ Time $\xrightarrow{have^{-1}}$ Post $\xrightarrow{write^{-1}}$ User
&Posts Containing Common Timestamps\\


$\Phi_7$
&U $\to$ P $\to$ L $\gets$ P $\gets$ U
&User $\xrightarrow{write}$ Post $\xrightarrow{have}$ Location $\xrightarrow{have^{-1}}$ Post $\xrightarrow{write^{-1}}$ User
&Posts Attaching Common Location Check-ins\\

\hline

\end{tabular}\vspace{-12pt}
}
\end{table*}

\subsection{Heterogeneous Network Meta Proximity}\label{subsec:heterogeneous_proximity}

For each attributed heterogeneous social network, the closeness among users can be denoted by the friendship links among them, where friends tend to be closer compared with user pairs without connections. Meanwhile, for the users who are not directly connected by the friendship links, few existing embedding methods can figure out their closeness, as these methods are mostly built based on the direct friendship link only. In this section, we propose to infer the potential closeness scores among the users with the heterogeneous information in the networks based on meta path concept \cite{SAH12}, which are formally called the \textit{meta proximity} in the paper.

\subsubsection{Friendship based Meta Proximity}\label{subsec:friendship_proximity}

In online social networks, the friendship links are the most obvious indicator of the social closeness among users. Online friends tend to be closer with each other compared with the user pairs who are not friends. Users' friendship links also carry important information about the local network structure information, which should be preserved in the embedding results. Based on such an intuition, we propose the \textit{friendship based meta proximity} concept as follows.

\noindent \textbf{Definition 3} (Friendship based Meta Proximity): For any two user nodes $u^{(1)}_i, u^{(1)}_j$ in an online social network (e.g., $G^{(1)}$), if $u^{(1)}_i$ and $u^{(1)}_j$ are friends in $G^{(1)}$, the \textit{friendship based meta proximity} between $u^{(1)}_i$ and $u^{(1)}_j$ in the network is $1$, otherwise the \textit{friendship based meta proximity} score between them will be $0$ instead. To be more specific, we can represent the \textit{friendship based meta proximity} score between users $u^{(1)}_i, u^{(1)}_j$ as $p^{(1)}(u^{(1)}_i,u^{(1)}_j) \in \{0, 1\}$, where $p^{(1)}(u^{(1)}_i,u^{(1)}_j) = 1$ iff $(u^{(1)}_i, u^{(1)}_j) \in \mathcal{E}^{(1)}_{u,u}$.

Based on the above definition, the \textit{friendship based meta proximity} scores among all the users in network $G^{(1)}$ can be represented as matrix $\mb{P}^{(1)}_{\Phi_0} \in \mathbb{R}^{|\mathcal{U}^{(1)}| \times |\mathcal{U}^{(1)}|}$, where entry ${P}^{(1)}_{\Phi_0}(i,j)$ equals to $p^{(1)}(u^{(1)}_i, u^{(1)}_j)$. Here $\Phi_0$ denotes the simplest meta path of length $1$ in the form $\mbox{U} \xrightarrow{\mbox{follow}} \mbox{U}$, and its formal definition will be introduced in the following subsection.

When network $G^{(1)}$ is an emerging online social network which has just started to provide services for a very short time, the friendship links among users in $G^{(1)}$ tend to be very limited (majority of the users are isolated in the network with few social connections). In other words, the \textit{friendship based meta proximity} matrix $\mb{P}^{(1)}_{\Phi_0}$ will be extremely sparse, where few entries will have value $1$ and most of the entries are $0$s. With such a sparse matrix, most existing embedding models will fail to work. The reason is that the sparse friendship information available in the network can hardly categorize the relative closeness relationships among the users (especially for those who are even not connected by friendship links), which renders these existing embedding models may project all the nodes to random regions.

To overcome such a problem, besides the social links, we propose to calculate the proximity scores for the users with the diverse link and attribute information in the heterogeneous networks in this paper. Based on a new concept named \textit{attribute augmented meta path}, a set of \textit{meta proximity} measures will be defined with each of the meta paths, which will be introduced in the following sections.

\subsubsection{Attribute Augmented Meta Path}\label{subsec:meta_path}

To handle the diverse links and attributes simultaneously in a unified analytic, we propose to treat the attributes as nodes as well and introduce the \textit{attribute augmented network}. If a node has certain attributes, a new type of link ``\textit{have}'' will be added to connected the node and the newly added attribute node. The structure of the \textit{attribute augmented network} can be described with the \textit{attribute augmented network schema} as follows.

\noindent \textbf{Definition 4} (Attribute Augmented Network Schema): Formally, the network schema of a given online social network $G^{(1)} = (\mathcal{V}, \mathcal{E})$ can be represented as $S_{G^{(1)}} = (\mathcal{N}_{\mathcal{V}} \cup \mathcal{N}_{\mathcal{T}}, \mathcal{R}_{\mathcal{E}} \cup \{\mbox{have}\})$, where $\mathcal{N}_{\mathcal{V}}$ and $\mathcal{N}_{\mathcal{T}}$ denote the set of node and attribute categories in the network, while  $\mathcal{R}_{\mathcal{E}}$ represents the set of link types in the network, and $\{\mbox{have}\}$ represents the relationship between node and attribute node types.

For instance, about the \textit{attributed heterogeneous social network} introduced after Definition 1 in Section~\ref{sec:formulation}, we can represent its network schema as $S_{G^{(1)}} = (\mathcal{N}_{\mathcal{V}} \cup \mathcal{N}_{\mathcal{T}}, \mathcal{R}_{\mathcal{E}} \cup \{\mbox{have}\})$. The node type set $\mathcal{N}_{\mathcal{V}}$ involves node types $\{\mbox{User}, \mbox{Post}\}$ (or $\{\mbox{U}, \mbox{P}\}$ for simplicity), while the attribute type set $\mathcal{N}_{\mathcal{T}}$ includes \{{Word}, {Time}, {Location}\} (or $\{\mbox{W}, \mbox{T}, \mbox{L}\}$ for short). As to the link types involved in the network, the link type set $\mathcal{R}_{\mathcal{E}}$ contains $\{\mbox{follow}, \mbox{write}\}$, which represents the friendship link type and the write link type respectively.

Based on the \textit{attribute augmented network schema}, we can represent the general correlation among users (especially those who are directly connected by friendship links) with the \textit{attributed augmented meta path} starting and ending with the user node type.

\noindent \textbf{Definition 5} (Attribute Augmented Meta Path): Given a network schema $S_{G^{(1)}}$, the \textit{attribute augmented meta path} denotes a sequence of node/attribute types connected by the link types or the ``$\mbox{have}$'' relation type (between node and attribute type). Formally, the \textit{attribute augmented meta path} (of length $k-1$, $k \ge 2$) can be represented as $\Phi: N_1 \xrightarrow{R_1} N_2 \xrightarrow{R_2} \cdots \xrightarrow{R_{k-1}} N_k$, where $N_1, \cdots, N_k \in \mathcal{N}_{\mathcal{V}} \cup \mathcal{N}_{\mathcal{T}}$ and $R_1, \cdots, R_{k-1} \in \mathcal{R}_{\mathcal{E}} \cup \mathcal{R}_{\mathcal{E}}^{-1} \cup \{\mbox{have}, \mbox{have}^{-1}\}$ (superscript $-1$ denotes the reverse of relation type direction). In the case that $N_1 = N_k = U$, i.e., meta paths starts and ends with the user node type, the meta paths will be called the \textit{social meta paths} specifically.

Based on the above definition, a set of different \textit{social meta path} $\{\Phi_0, \Phi_1, \Phi_2, \cdots, \Phi_7\}$ can be extracted from the network, whose notations, concrete representations and the physical meanings are illustrated in Table~\ref{tab:meta_path}. Here, meta paths $\Phi_0-\Phi_4$ are all based on the user node type and follow link type; meta paths $\Phi_5-\Phi_7$ involve the user, post node type, attribute node type, as well as the \textit{write} and \textit{have} link type. Based on each of the meta paths, there will exist a set of concrete meta path instances connecting users in the networks. For instance, given a user pair $u$ and $v$, they may have been checked-in at 5 different common locations, which will introduce $5$ concrete meta path instance of meta path $\Phi_7$ connecting $u$ and $v$ indicating their strong closeness (in location check-ins). In the next subsection, we will introduce how to calculate the proximity score for the users based on these extracted meta paths.

\subsubsection{Heterogeneous Network Meta Proximity}\label{subsec:meta_path_proximity}

The set of \textit{attribute augmented social meta paths} $\{\Phi_0, \Phi_1, \Phi_2, \cdots, \Phi_7\}$ extracted in the previous subsection create different kinds of correlations among users (especially for those who are not directed connected by friendship links). With these \textit{social meta paths}, different types of proximity scores among the users can be captured. For instance, for the users who are not friends but share lots of common friends, they may also know each other and can be close to each other; for the users who frequently checked-in at the same places, they tend to be more close to each other compared with those isolated ones with nothing in common. Therefore, these meta paths can help capture much broader network structure information compared with the local structure captured by the \textit{friendship based meta proximity} covered in Section~\ref{subsec:friendship_proximity}. In this part, we will introduce the method to calculate the proximity scores among users based on these \textit{social meta paths}.

As shown in Table~\ref{tab:meta_path}, all the social meta paths extracted from the networks can be represented as set $\{\Phi_0, \Phi_1, \cdots, \Phi_7\}$. Given a pair of users, e.g., $u^{(1)}_i$ and $u^{(1)}_j$, based on meta path $\Phi_k \in \{\Phi_0, \Phi_1, \cdots, \Phi_7\}$, we can represent the set of meta path instances connecting $u^{(1)}_i$ and $u^{(1)}_j$ as $\mathcal{P}_{\Phi_k}^{(1)}(u^{(1)}_i, u^{(1)}_j)$. Users $u^{(1)}_i$ and $u^{(1)}_j$ can have multiple meta path instances going into/out from them. Formally, we can represent all the meta path instances going out from user $u^{(1)}_i$ (or going into $u^{(1)}_j$), based on meta path $\Phi_k$, as set $\mathcal{P}_{\Phi_k}^{(1)}(u^{(1)}_i, \cdot)$ (or $\mathcal{P}_{\Phi_k}^{(1)}(\cdot, u^{(1)}_j)$). The proximity score between $u^{(1)}_i$ and $u^{(1)}_j$ based on meta path $\Phi_k$ can be represented as the following \textit{meta proximity} concept formally.

\noindent \textbf{Definition 6} (Meta Proximity): Based on meta path $\Phi_k$, the meta proximity between users $u^{(1)}_i$ and $u^{(1)}_j$ in $G^{(1)}$ can be represented as 
$$p^{(1)}_{\Phi_k}(u^{(1)}_i, u^{(1)}_j) = \frac{2|\mathcal{P}_{\Phi_k}^{(1)}(u^{(1)}_i, u^{(1)}_j)|}{|\mathcal{P}_{\Phi_k}^{(1)}(u^{(1)}_i, \cdot)| + |\mathcal{P}_{\Phi_k}^{(1)}(\cdot, u^{(1)}_j)|}.$$

\textit{Meta proximity} considers not only the meta path instances between users but also penalizes the number of meta path instances going out from/into $u^{(1)}_i$ and $u^{(1)}_j$ at the same time. It is also reasonable. For instance, sharing some common location check-ins with some extremely active users (who have thousands of checkins) may not necessarily indicate closeness with them, since they may have common check-ins with almost all other users simply due to his very large check-in record volume instead of their closeness.

With the above meta proximity definition, we can represent the meta proximity scores among all users in the network $G^{(1)}$ based on meta path $\Phi_k$ as matrix $\mb{P}^{(1)}_{\Phi_k} \in \mathbb{R}^{|\mathcal{U}^{(1)}| \times |\mathcal{U}^{(1)}|}$, where entry ${P}^{(1)}_{\Phi_k}(i,j) = p^{(1)}_{\Phi_k}(u^{(1)}_i, u^{(1)}_j)$. All the meta proximity matrices defined for network $G^{(1)}$ can be represented as $\{\mb{P}^{(1)}_{\Phi_k}\}_{\Phi_k}$. Based on the meta paths extracted for network $G^{(2)}$, similar matrices can be defined as well, which can be denoted as $\{\mb{P}^{(2)}_{\Phi_k}\}_{\Phi_k}$.

\begin{figure*}
	\vspace{-30pt}
	\centering
	\includegraphics[width=0.8\textwidth]{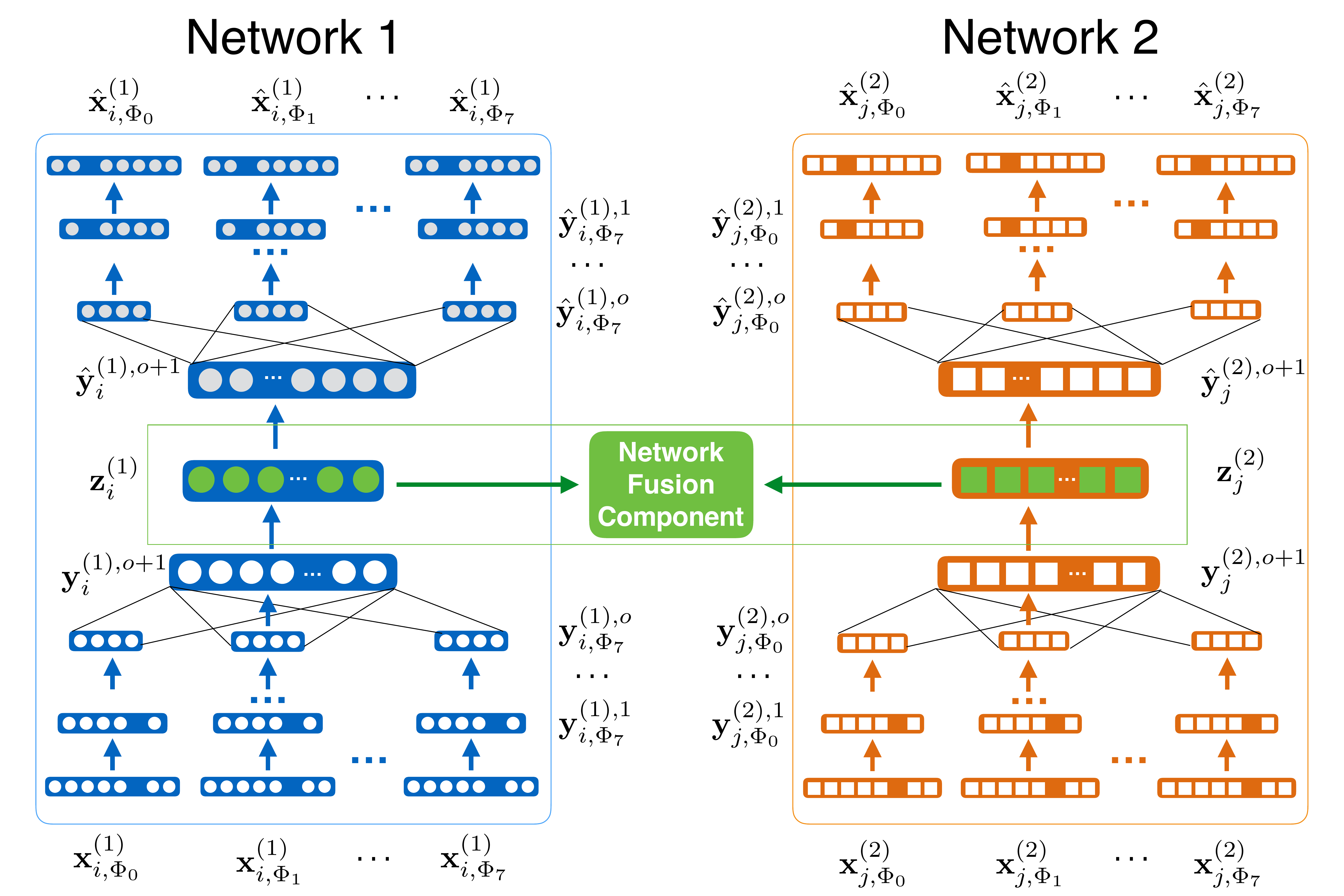}
	\vspace{-10pt}
	\caption{The {\our} Framework.}
	\label{fig:deep}\vspace{-15pt}
\end{figure*}

\subsection{Deep Network Synergistic Embedding}\label{subsec:deep}

With these calculated \textit{meta proximity} introduced in the previous section, we will introduce the embedding framework {\our} in this part. {\our} is based on the \textit{aligned auto-encoder model}, which extends the traditional \textit{deep auto-encoder model} to the \textit{multiple aligned heterogeneous networks} scenario. To make this paper self-contained, we will first briefly introduce some background knowledge about the auto-encoder model first in Section~\ref{subsec:auto_encoder}. After that, we will talk about the embedding model component for one single heterogeneous network in Section~\ref{subsec:single}, which takes the various meta proximity matrices as input. {\our} effectively couples the embedding process of the emerging network with other aligned mature networks, where cross-network information exchange and refinement is achieved via the loss term defined based on the anchor links.

\subsubsection{Deep Auto-Encoder Model Review}\label{subsec:auto_encoder}

Auto-encoder is an unsupervised neural network model, which projects the instances (in original feature representations) into a lower-dimensional feature space via a series of non-linear mappings. Auto-encoder model involves two steps: encoder and decoder. The encoder part projects the original feature vectors to the objective feature space, while the decoder step recovers the latent feature representation to a reconstruction space. In auto-encoder model, we generally need to ensure that the original feature representation of instances should be as similar to the reconstructed feature representation as possible.

Formally, let $\mb{x}_i$ represent the original feature representation of instance $i$, and $\mb{y}^1_i, \mb{y}^2_i, \cdots, \mb{y}^o_i$ be the latent feature representation of the instance at hidden layers $1, 2, \cdots, o$ in the encoder step, the encoding result in the objective feature space can be represented as $\mb{z}_i \in \mathbb{R}^{d}$ with dimension $d$. Formally, the relationship between these variables can be represented with the following equations:
$$\begin{cases}
\mb{y}^1_i &= \sigma (\mb{W}^1 \mb{x}_i + \mb{b}^1),\\
\mb{y}^k_i &= \sigma (\mb{W}^k \mb{y}^{k-1}_i + \mb{b}^k), \forall k \in \{2, 3, \cdots, o\},\\
\mb{z}_i &= \sigma (\mb{W}^{o+1} \mb{y}^o_i + \mb{b}^{o+1}).
\end{cases}$$

Meanwhile, in the decoder step, the input will be the latent feature vector $\mb{z}_i$ (i.e., the output of the encoder step), and the final output will be the reconstructed vector $\hat{\mb{x}}_i$. The latent feature vectors at each hidden layers can be represented as $\hat{\mb{y}}^{o}_i, \hat{\mb{y}}^{o-1}_i, \cdots, \hat{\mb{y}}^{1}_i$. The relationship between these vector variables can be denoted as
$$\begin{cases}
\hat{\mb{y}}^o_i &= \sigma (\hat{\mb{W}}^{o+1} \mb{z}_i + \hat{\mb{b}}^{o+1}),\\
\hat{\mb{y}}^{k-1}_i &= \sigma (\hat{\mb{W}}^k \hat{\mb{y}}^{k}_i + \hat{\mb{b}}^k), \forall k \in \{2, 3, \cdots, o\},\\
\hat{\mb{x}}_i &= \sigma(\hat{\mb{W}}^1 \hat{\mb{y}}^{1}_i + \hat{\mb{b}}^1).
\end{cases}$$

The objective of the auto-encoder model is to minimize the loss between the original feature vector $\mb{x}_i$ and the reconstructed feature vector $\hat{\mb{x}}_i$ of all the instances in the network. Formally, the loss term can be represented as
$$\mathcal{L} = \sum_{i} \left \| {\mb{x}}_i - \hat{\mb{x}}_i\right\|_2^2.$$

\subsubsection{Deep {\oursingle} Model}\label{subsec:single}

When applying the auto-encoder model for one single homogeneous network, e.g., for $G^{(1)}$, we can fit the model with the node meta proximity feature vectors, i.e., rows corresponding to users in matrix $\mb{P}^{(1)}_{\Phi_0}$ (introduced in Section~\ref{subsec:friendship_proximity}). In the case that $G^{(1)}$ is heterogeneous, multiple node \textit{meta proximity} matrices have been defined before (i.e., $\{\mb{P}^{(1)}_{\Phi_0}, \mb{P}^{(1)}_{\Phi_1}, \cdots, \mb{P}^{(1)}_{\Phi_7}\}$), how to fit these matrices simultaneously to the auto-encoder models is an open problem. In this part, we will introduce the single-heterogeneous-network version of framework {\our}, namely {\oursingle}, which will be used as an important component of framework {\our} as well. For each user node in the network, {\oursingle} computes the embedding vector based on each of the proximity matrix independently first, which will be further fused to compute the final latent feature vector in the output hidden layer.

As shown in the architecture in Figure~\ref{fig:deep} (either the left component for network 1 or the right component for network 2), about the same instance, {\oursingle} takes different feature vectors extracted from the meta paths $\{\Phi_0, \Phi_1, \cdots, \Phi_7\}$ as the input. For each meta path, a series of separated encoder and decoder steps are carried out simultaneously, whose latent vectors are fused together to calculate the final embedding vector $\mb{z}^{(1)}_i \in \mathbb{R}^{d^{(1)}}$ for user $u_i^{(1)} \in \mathcal{V}^{(1)}$. In the {\oursingle} model, the input feature vectors (based on meta path $\Phi_k \in \{\Phi_0, \Phi_1, \cdots, \Phi_7\}$) of user $u_i$ can be represented as $\mb{x}^{(1)}_{i, \Phi_k}$, which denotes the row corresponding to users $u_i^{(1)}$ in matrix $\mb{P}^{(1)}_{\Phi_k}$ defined before. Meanwhile, the latent representation of the instance based on the feature vector extracted via meta path $\Phi_k$ at different hidden layers can be represented as $\{\mb{y}^{(1),1}_{i, \Phi_k}, \mb{y}^{(1),2}_{i, \Phi_k}, \cdots, \mb{y}^{(1),o}_{i, \Phi_k}\}$. 


One of the significant difference of model {\oursingle} from traditional auto-encoder model lies in the (1) combination of multiple hidden vectors $\{\mb{y}^{(1),o}_{i, \Phi_0}, \mb{y}^{(1),o}_{i, \Phi_1}, \cdots, \mb{y}^{(1),o}_{i, \Phi_7}\}$ to obtain the embedding vector $\mb{z}^{(1)}_i$ in the encoder step, and (2) the dispatch of embedding vector $\mb{z}^{(1)}_i$ back to the hidden vectors in the decoder step. As shown in the architecture, formally, these extra steps can be represented as
\begin{align*}
\begin{cases}
&\mbox{\# extra encoder steps}\\
&\mb{y}^{(1), o+1}_i =  \sigma(\sum_{\Phi_k \in \{\Phi_0, \cdots, \Phi_7\}} \mb{W}^{(1),o+1}_{\Phi_k} \mb{y}^{(1),o}_{i, \Phi_k} + \mb{b}^{(1),o+1}_{\Phi_k}),\\
&\mb{z}^{(1)}_i = \sigma(\mb{W}^{(1),o+2} \mb{y}^{(1), o+1}_i + \mb{b}^{(1),o+2}).\\
&\mbox{\# extra decoder steps }\\
&\hat{\mb{y}}^{(1), o+1}_i =  \sigma(\hat{\mb{W}}^{(1),o+2} \mb{z}^{(1)}_i + \hat{\mb{b}}^{(1),o+2}),\\
&\hat{\mb{y}}^{(1),o}_{i, \Phi_k} = \sigma(\hat{\mb{W}}^{(1),o+1}_{\Phi_k} \hat{\mb{y}}^{(1), o+1}_i + \hat{\mb{b}}^{(1),o+1}_{\Phi_k}).
\end{cases}
\end{align*}
In the fusion and dispatch steps, full connection layers are used in order to incorporate all the information captured by all the meta paths.

What's more, since the input feature vectors are extremely sparse (lots of the entries are $0$s), simply feeding them to the model may lead to some trivial solutions, like $\mb{0}$ vectors for both $\mb{z}^{(1)}_i$ and the decoded vectors $\hat{\mb{x}}_{i, \Phi_k}^{(1)}$. To overcome such a problem, another significant difference of model {\oursingle} from traditional auto-encoder model lies in the loss function definition, where the loss introduced by the non-zero features will be assigned with a larger weight. In addition, by adding the loss function for each of the meta paths, the final loss function in {\oursingle} can be formally represented as
$$\mathcal{L}^{(1)} = \sum_{\Phi_k \in \{\Phi_0, \cdots, \Phi_7\}}\sum_{u_i \in \mathcal{V}} \left \| \left( \mb{x}^{(1)}_{i, \Phi_k} - \hat{\mb{x}}^{(1)}_{i,\Phi_k} \right) \odot \mb{b}^{(1)}_{i,\Phi_k} \right\|_2^2,$$
where vector $\mb{b}^{(1)}_{i,\Phi_k}$ is the weight vector corresponding to feature vector $\mb{x}^{(1)}_{i,\Phi_k}$. Entries in vector $\mb{b}^{(1)}_{i,\Phi_k}$ are filled with value $1$s except the entries corresponding to non-zero element in $\mb{x}^{(1)}_{i, \Phi_k}$, which will be assigned with value $\gamma$ ($\gamma > 1$ denoting a larger weight to fit these features). Here, we need to add a remark that ``simply discarding the entries corresponding zero values in the input vectors from the loss function'' will not work here, since it will allow the model to decode there entries to any random values, which will not be what we want. In a similar way, we can define the loss function for the embedding result in network $G^{(2)}$, which can be formally represented as $\mathcal{L}^{(2)}$.

\subsubsection{Deep {\our} Framework}\label{subsec:framework}

{\oursingle} has incorporate all these heterogeneous information in the model building, the meta proximity calculated based on which can help differentiate the closeness among different users. However, for the emerging networks which just start to provide services, the information sparsity problem may affect the performance of {\oursingle} significantly. In this part, we will introduce {\our}, which couples the embedding process of the emerging network with another mature aligned network. By accommodating the embedding between the aligned networks, information can be transferred from the aligned mature network to refine the embedding results in the emerging network effectively. The complete architecture of {\our} is shown in Figure~\ref{fig:deep}, which involve the {\oursingle} components for each of the aligned networks, where the information transfer component aligns these separated {\oursingle} models together.

To be more specific, given a pair of aligned heterogeneous networks $\mathcal{G} = ((G^{(1)}, G^{(2)}), \mathcal{A}^{(1,2)})$ ($G^{(1)}$ is an emerging network and $G^{(2)}$ is a mature network), we can represent the embedding results as matrices  $\mb{Z}^{(1)} \in \mathbb{R}^{|\mathcal{U}^{(1)}| \times d^{(1)}}$ and $\mb{Z}^{(2)} \in \mathbb{R}^{|\mathcal{U}^{(2)}| \times d^{(2)}}$ for all the user nodes in $G^{(1)}$ and $G^{(2)}$ respectively. The $i_{th}$ row of matrix $\mb{Z}^{(1)}$ (or the $j_{th}$ row of matrix $\mb{Z}^{(2)}$) denotes the encoded feature vector of user $u^{(1)}_i$ in $G^{(1)}$ (or $u^{(2)}_j$ in $G^{(2)}$). If $u^{(1)}_i$ and $u^{(2)}_j$ are the same user, i.e., $(u^{(1)}_i, u^{(2)}_j) \in \mathcal{A}^{(1,2)}$, by placing vectors $\mb{Z}^{(1)}(i,:)$ and $\mb{Z}^{(2)}(j,:)$ in a close region in the embedding space, we can use the information from $G^{(2)}$ to refine the embedding result in $G^{(1)}$.

Information transfer is achieved based on the anchor links, and we only care about the anchor users. To adjust the rows of matrices $\mb{Z}^{(1)}$ and $\mb{Z}^{(2)}$ to remove non-anchor users and make the same rows correspond to the same user, we introduce the binary inter-network transitional matrix $\mb{T}^{(1,2)} \in \mathbb{R}^{|\mathcal{U}^{(1)}| \times |\mathcal{U}^{(2)}|}$. Entry $T^{(1,2)}(i,j) = 1$ iff the corresponding users are connected by anchor links, i.e., $(u^{(1)}_i, u^{(2)}_j) \in \mathcal{A}^{(1,2)}$. Furthermore, the encoded feature vectors for users in these two networks can be of different dimensions, i.e., $d^{(1)} \neq d^{(2)}$, which can be accommodated via the projection $\mb{W}^{(1,2)} \in \mathbb{R}^{d^{(1)} \times d^{(2)}}$. 

Formally, the introduced \textit{information fusion loss} between networks $G^{(1)}$ and $G^{(2)}$ can be represented as
$$\mathcal{L}^{(1,2)} = \left\| (\mb{T}^{(1,2)})^\top \mb{Z}^{(1)} \mb{W}^{(1,2)} - \mb{Z}^{(2)}  \right\|_F^2.$$
By minimizing the \textit{information fusion loss} function $\mathcal{L}^{(1,2)}$, we can use the anchor users' embedding vectors from the mature network $G^{(2)}$ to adjust his embedding vectors in the emerging network $G^{(1)}$. Even through in such a process the embedding vector in $G^{(2)}$ can be undermined by $G^{(1)}$, it will not be a problem since $G^{(1)}$ is our target network and we only care about the embedding result of the emerging network $G^{(1)}$ in the paper. 

\begin{table}[t]
\vspace{-30pt}
\caption{Properties of the Heterogeneous Networks}
\label{tab:datastat}
\centering
\begin{tabular}{clrr}
\toprule
&&\multicolumn{2}{c}{network}\\
\cmidrule{3-4}
&property &\textbf{Twitter} &\textbf{Foursquare}   \\
\midrule 
\multirow{3}{*}{\# node}
&user   & 5,223 & 5,392 \\
&tweet/tip  & 9,490,707 & 48,756 \\
&location & 297,182 & 38,921 \\
\midrule 
\multirow{3}{*}{\# link}
&friend/follow    &164,920  &76,972 \\
&write    & 9,490,707 & 48,756 \\
&locate   & 615,515 & 48,756 \\
\bottomrule
\end{tabular}\vspace{-15pt}
\end{table}

The complete objective function of framework include the loss terms introduced by the component {\oursingle} for networks $G^{(1)}$, $G^{(2)}$, and the \textit{information fusion loss}, which can be denoted as
$$\mathcal{L}(G^{(1)}, G^{(2)}) = \mathcal{L}^{(1)} + \mathcal{L}^{(2)} + \alpha \cdot \mathcal{L}^{(1,2)}  + \beta \cdot \mathcal{L}_{reg}.$$
Parameters $\alpha$ and $\beta$ denote the weights of the \textit{information fusion loss} term and the regularization term. In the objective function, term $\mathcal{L}_{reg}$ is added to the above objective function to avoid overfitting, which can be formally represented as
\begin{align*}
\begin{cases}
&\mathcal{L}_{reg} = \mathcal{L}_{reg}^{(1)} + \mathcal{L}_{reg}^{(2)} + \mathcal{L}_{reg}^{(1,2)},\\
&\mathcal{L}_{reg}^{(1)} = \sum_{i}^{o^{(1)}+2} \sum_{\Phi_k \in \{\Phi_0, \cdots, \Phi_7\}} \left( \left\| \mb{W}^{(1),i}_{\Phi_k} \right\|_F^2 + \left\| \hat{\mb{W}}^{(1),i}_{\Phi_k} \right\|_F^2 \right),\\
&\mathcal{L}_{reg}^{(2)} = \sum_{i}^{o^{(2)}+2} \sum_{\Phi_k \in \{\Phi_0, \cdots, \Phi_7\}} \left( \left\| \mb{W}^{(2),i}_{\Phi_k} \right\|_F^2 + \left\| \hat{\mb{W}}^{(2),i}_{\Phi_k} \right\|_F^2 \right),\\
&\mathcal{L}_{reg}^{(1,2)} =  \left\| \mb{W}^{(1,2)} \right\|_F^2.
\end{cases}
\end{align*}

To optimize the above objective function, we utilize Stochastic Gradient Descent (SGD). To be more specific, the training process involves multiple epochs. In each epoch, the training data is shuffled and a minibatch of the instances are sampled to update the parameters with SGD. Such a process continues until either convergence or the training epochs have been finished.


\begin{table*}[t]
\vspace{-30pt}
\caption{Link prediction result of the comparison methods (parameter $\lambda$ changes in $\{10\%, 20\%, \cdots, 100\%\}$, $\theta = 1$).}
\label{tab:link_prediction_result}
\centering
{\tiny
\begin{tabular}{lrcccccccccc}
\toprule
\multicolumn{2}{l}{ }&\multicolumn{10}{c}{Sampling Ratio $\lambda$}\\
\cmidrule{3-12}
metric &method &$10\%$ &$20\%$  &$30\%$   &$40\%$   &$50\%$   &$60\%$   &$70\%$   &$80\%$    &$90\%$  &$100\%$ \\
\midrule
\multirow{5}{*}{\rotatebox{90}{AUC}}
&{\our} &\textbf{0.792}$\pm$\textbf{0.007}  &\textbf{0.822}$\pm$\textbf{0.006}  &\textbf{0.838}$\pm$\textbf{0.005}  &\textbf{0.843}$\pm$\textbf{0.003}  &\textbf{0.847}$\pm$\textbf{0.003}  &\textbf{0.850}$\pm$\textbf{0.003}  &\textbf{0.852}$\pm$\textbf{0.002}  &\textbf{0.850}$\pm$\textbf{0.004}  &\textbf{0.852}$\pm$\textbf{0.003}  &\textbf{0.852}$\pm$\textbf{0.004}  \\
&{\oursingle} &0.774$\pm$0.006 &0.795$\pm$0.005  &0.802$\pm$0.006  &0.809$\pm$0.004   &0.815$\pm$0.005 &0.822$\pm$0.005  &0.827$\pm$0.006  &0.826$\pm$0.005  &0.830$\pm$0.004  &0.833$\pm$0.003
\\
\cmidrule{2-12}
&Auto-encoder&0.697$\pm$0.006 &0.731$\pm$0.006  &0.752$\pm$0.005  &0.761$\pm$0.005  &0.761$\pm$0.005  &0.763$\pm$0.004  &0.763$\pm$0.004 &0.770$\pm$0.003 &0.773$\pm$0.005 &0.777$\pm$0.005
\\
&LINE&0.694$\pm$0.008 &0.716$\pm$0.003  &0.731$\pm$0.007  &0.738$\pm$0.005  &0.741$\pm$0.006  &0.744$\pm$0.005  &0.748$\pm$0.005 &0.748$\pm$0.008 &0.750$\pm$.003 &0.750$\pm$0.006
\\
&DeepWalk&0.671$\pm$0.010 &0.661$\pm$0.011  &0.670$\pm$0.009  &0.675$\pm$0.009  &0.682$\pm$0.005  &0.687$\pm$0.004  &0.701$\pm$0.007 &0.718$\pm$0.007 &0.733$\pm$0.008  &0.747$\pm$0.006  \\
\cmidrule{1-12}
\multirow{5}{*}{\rotatebox{90}{Accuracy}}
&{\our} &\textbf{0.719}$\pm$\textbf{0.006}  &\textbf{0.748}$\pm$\textbf{0.005}  &\textbf{0.763}$\pm$\textbf{0.004}  &\textbf{0.767}$\pm$\textbf{0.003}  &\textbf{0.773}$\pm$\textbf{0.003}  &\textbf{0.775}$\pm$\textbf{0.004}  &\textbf{0.777}$\pm$\textbf{0.003} &\textbf{0.775}$\pm$\textbf{0.003}  &\textbf{0.777}$\pm$\textbf{0.004} &\textbf{0.777}$\pm$\textbf{0.004}  \\
&{\oursingle} &0.704$\pm$0.007 &0.723$\pm$0.004  &0.728$\pm$0.006  &0.737$\pm$0.003  &0.739$\pm$0.006  &0.747$\pm$0.005  &0.753$\pm$0.006   &0.754$\pm$0.006  &0.757$\pm$0.005 &0.761$\pm$0.003\\
\cmidrule{2-12}
&Auto-encoder&0.642$\pm$0.005 &0.668$\pm$0.005  &0.684$\pm$0.005  &0.692$\pm$0.005  &0.691$\pm$0.005  &0.691$\pm$0.004  &0.691$\pm$0.004   &0.699$\pm$0.004  &0.700$\pm$0.005 &0.703$\pm$0.005  \\
&LINE&0.637$\pm$0.005 &0.666$\pm$0.004  &0.676$\pm$0.008  &0.676$\pm$0.005  &0.677$\pm$0.004  &0.679$\pm$0.006  &0.679$\pm$0.005   &0.679$\pm$0.008    &0.681$\pm$0.003 &0.682$\pm$0.007  \\
&DeepWalk&0.632$\pm$0.008 &0.626$\pm$0.009  &0.633$\pm$0.008  &0.634$\pm$0.008  &0.637$\pm$0.006  &0.641$\pm$0.004  &0.655$\pm$0.007 &0.669$\pm$0.006 &0.680$\pm$0.006  &0.687$\pm$0.004  \\
\cmidrule{1-12}
\multirow{5}{*}{\rotatebox{90}{Recall}}
&{\our} &0.641$\pm$0.008  &0.702$\pm$0.011  &0.732$\pm$0.007  &0.746$\pm$0.008  &\textbf{0.755}$\pm$\textbf{0.008}  &\textbf{0.761}$\pm$\textbf{0.006}  &\textbf{0.767}$\pm$\textbf{0.006} &\textbf{0.768}$\pm$\textbf{0.003}  &\textbf{0.771}$\pm$\textbf{0.005}   &\textbf{0.772}$\pm$\textbf{0.008}  \\
&{\oursingle} &0.641$\pm$0.011 &0.689$\pm$0.006  &0.692$\pm$0.010  &0.703$\pm$0.009  &0.707$\pm$0.009  &0.713$\pm$0.013  &0.720$\pm$0.010  &0.719$\pm$0.012  &0.725$\pm$0.009  &0.731$\pm$0.008\\
\cmidrule{2-12}
&Auto-encoder&0.564$\pm$0.016 &0.649$\pm$0.016  &0.714$\pm$0.007  &0.743$\pm$0.013  &0.726$\pm$0.012  &0.680$\pm$0.009  &0.671$\pm$0.007  &0.681$\pm$0.008  &0.680$\pm$0.008  &0.687$\pm$0.007  \\
&LINE&\textbf{0.819}$\pm$\textbf{0.008} &\textbf{0.770}$\pm$\textbf{0.007}  &\textbf{0.800}$\pm$\textbf{0.008}  &\textbf{0.749}$\pm$\textbf{0.011}  &0.740$\pm$0.008  &0.731$\pm$0.010  &0.724$\pm$0.007  &0.721$\pm$0.007  &0.715$\pm$0.005  &0.716$\pm$0.009  \\
&DeepWalk&0.645$\pm$0.020 &0.658$\pm$0.020  &0.678$\pm$0.016  &0.681$\pm$0.016  &0.680$\pm$0.016  &0.682$\pm$0.010  &0.692$\pm$0.009 &0.702$\pm$0.008 &0.706$\pm$0.005  &0.707$\pm$0.007  \\
\cmidrule{1-12}
\multirow{5}{*}{\rotatebox{90}{F1}}
&{\our} &\textbf{0.700}$\pm$\textbf{0.007}  &\textbf{0.735}$\pm$\textbf{0.008}  &\textbf{0.756}$\pm$\textbf{0.006}  &\textbf{0.762}$\pm$\textbf{0.006}  &\textbf{0.769}$\pm$\textbf{0.005}  &\textbf{0.772}$\pm$\textbf{0.005}  &\textbf{0.775}$\pm$\textbf{0.004} &\textbf{0.774}$\pm$\textbf{0.004}  &\textbf{0.776}$\pm$\textbf{0.005} &\textbf{0.776}$\pm$\textbf{0.006}  \\
&{\oursingle} &0.684$\pm$0.009 &0.711$\pm$0.005  &0.718$\pm$0.008  &0.728$\pm$0.006  &0.731$\pm$0.007  &0.738$\pm$0.008  &0.744$\pm$0.007  &0.745$\pm$0.009   &0.749$\pm$0.006 &0.753$\pm$0.005\\
\cmidrule{2-12}
&Auto-encoder&0.612$\pm$0.009 &0.662$\pm$0.009  &0.693$\pm$0.007  &0.707$\pm$0.007  &0.701$\pm$0.007  &0.688$\pm$0.006  &0.685$\pm$0.006   &0.694$\pm$0.006 &0.694$\pm$0.007  &0.698$\pm$0.008  \\
&LINE&0.693$\pm$0.006 &0.698$\pm$0.005  &0.712$\pm$0.008  &0.698$\pm$0.007  &0.696$\pm$0.006  &0.695$\pm$0.009  &0.693$\pm$0.006  &0.692$\pm$0.009     &0.692$\pm$0.005 &0.693$\pm$0.009  \\
&DeepWalk&0.636$\pm$0.011 &0.637$\pm$0.014  &0.648$\pm$0.010  &0.650$\pm$0.010  &0.652$\pm$0.010  &0.655$\pm$0.006  &0.667$\pm$0.008    &0.679$\pm$0.006   &0.688$\pm$0.006 &0.693$\pm$0.005  \\
\bottomrule

\end{tabular}\vspace{-15pt}
}
\end{table*}

\section{Experiments}\label{sec:experiment}

To demonstrate the effectiveness of the learnt embedding feature vectors, extensive experiments have been done on real-world aligned heterogeneous social networks, Foursquare and Twitter. Two different tasks are done in this section for embedding result evaluation purposes, which include \textit{link prediction} and \textit{community detection}. In this section, we will provide some basic descriptions about the aligned heterogeneous social network dataset first. After that, we will introduce the experimental settings, covering the comparison embedding methods, as well as the experimental settings, and evaluation metrics for \textit{link prediction} and \textit{community detection} tasks. Finally, we will show the experimental results about \textit{link prediction} and \textit{community detection}, followed by the parameter analysis.

\subsection{Dataset Description}

The data used in the experiments include two \textit{aligned heterogeneous social networks} Foursquare and Twitter simultaneously. The basic statistical information about the Foursquare and Twitter datasets is available in Table~\ref{tab:datastat}. The data crawling strategy and method is introduced in great deital in \cite{KZY13, ZYZ14}.

\begin{itemize}

\item \textit{Twitter}: Twitter is a famous \textit{micro-blogging site} that allows users to write, read and share posts with their friends online. We have crawled $5,223$ Twitter users, and $164,920$ follow links among them. These crawled Twitter users have posted $9,490,707$ tweets, among which $615,515$ have location checkins.

\item \textit{Foursquare}: Foursquare is a famous \textit{location based social network} (LBSN), which provides users with various kinds of location-related services. From Foursquare, we have crawled $5,392$ users together with $76,972$ friendship links among them. These Foursquare users have written $48,756$ posts which all attach location checkins. Among these $5,392$ crawled Foursquare users, $3,388$ of them are aligned by anchor links with Twitter. 

\end{itemize}

In the experiments, we will use Foursquare as the emerging network and Twitter as the aligned mature network, since Twitter has more dense information than Foursquare. The results for the reverse case (Foursquare: mature; Twitter: emerging) are not shown here due to the limited space.\\

\noindent \textbf{Source Code}: The source code of {\our} is available at site: http://www.ifmlab.org/files/code/Aligned-Autoencoder.zip.

\subsection{Experimental Settings}

In this paper, we are mainly focused on studying the embedding models, and different network embedding comparison methods will be introduced first. After that, we will introduce the experimental settings for both \textit{link prediction} and \textit{community detection} tasks, which will be used as the evaluation tasks to determine whether the embedding results are good or not. A set of frequently used evaluation metrics for link prediciton and community detection will be introduced afterwards. 

\subsubsection{Embedding Comparison Methods}

\begin{table*}[t]
\vspace{-30pt}

\caption{Community detection result of the comparison methods (parameter $\lambda$ changes in $\{10\%, 20\%, \cdots, 100\%\}$, $k = 10$).}
\label{tab:community_detection_result}
\centering
{\tiny
\begin{tabular}{lrcccccccccc}
\toprule
\multicolumn{2}{l}{ }&\multicolumn{10}{c}{Sampling Ratio $\lambda$}\\
\cmidrule{3-12}
metric &method &$10\%$ &$20\%$  &$30\%$   &$40\%$   &$50\%$   &$60\%$   &$70\%$   &$80\%$    &$90\%$  &$100\%$\\
\midrule
\multirow{5}{*}{\rotatebox{90}{Density}}
&{\our} &\textbf{0.002}$\pm$\textbf{0.000}  &\textbf{0.003}$\pm$\textbf{0.000}  &\textbf{0.003}$\pm$\textbf{0.000}  &\textbf{0.003}$\pm$\textbf{0.000}  &\textbf{0.004}$\pm$\textbf{0.000}  &\textbf{0.003}$\pm$\textbf{0.000}  &\textbf{0.004}$\pm$\textbf{0.000}  &\textbf{0.003}$\pm$\textbf{0.000}  &\textbf{0.003}$\pm$\textbf{0.000}  &\textbf{0.004}$\pm$\textbf{0.000}  \\
&{\oursingle} &\textbf{0.002}$\pm$\textbf{0.000} &0.002$\pm$0.000  &\textbf{0.003}$\pm$\textbf{0.000} &\textbf{0.003}$\pm$\textbf{0.000}  &0.003$\pm$0.000  &\textbf{0.003}$\pm$\textbf{0.000}  &0.003$\pm$0.000  &\textbf{0.003}$\pm$\textbf{0.000}  &\textbf{0.003}$\pm$\textbf{0.000}  &0.003$\pm$0.000 \\
\cmidrule{2-12}
&Auto-encoder&\textbf{0.002}$\pm$\textbf{0.000} &0.002$\pm$0.000  &0.002$\pm$0.000  &0.002$\pm$0.000  &0.003$\pm$0.000  &\textbf{0.003}$\pm$\textbf{0.000}  &0.003$\pm$0.000  &\textbf{0.003}$\pm$\textbf{0.000}  &\textbf{0.003}$\pm$\textbf{0.000}  &0.003$\pm$0.000    \\
&LINE&0.001$\pm$0.000  &0.002$\pm$0.000  &0.002$\pm$0.000  &0.002$\pm$0.000  &0.002$\pm$0.000  &0.002$\pm$0.000  &0.002$\pm$0.000  &0.002$\pm$0.000  &0.002$\pm$0.000  &0.002$\pm$0.000   \\
&DeepWalk&0.001$\pm$0.000  &0.001$\pm$0.000  &0.001$\pm$0.000  &0.001$\pm$0.000  &0.001$\pm$0.000  &0.001$\pm$0.000  &0.001$\pm$0.000  &0.001$\pm$0.000  &0.001$\pm$0.000  &0.001$\pm$0.000 \\
\cmidrule{1-12}
\multirow{5}{*}{\rotatebox{90}{Separability}}
&{\our} &0.230$\pm$0.007 &\textbf{0.296}$\pm$\textbf{0.027}  &\textbf{0.360}$\pm$\textbf{0.021}  &\textbf{0.369}$\pm$\textbf{0.028}  &\textbf{0.440}$\pm$\textbf{0.023}  &\textbf{0.381}$\pm$\textbf{0.028}  &\textbf{0.466}$\pm$\textbf{0.022}  &0.347$\pm$0.004  &\textbf{0.398}$\pm$\textbf{0.025}  &\textbf{0.414}$\pm$\textbf{0.018}\\
&{\oursingle} &\textbf{0.251}$\pm$\textbf{0.033}  &0.235$\pm$0.007  &0.288$\pm$0.013  &0.294$\pm$0.010  &0.311$\pm$0.020  &0.282$\pm$0.018  &0.298$\pm$0.007  &\textbf{0.370}$\pm$\textbf{0.011}  &0.362$\pm$0.010  &0.372$\pm$0.017\\
\cmidrule{2-12}
&Auto-encoder&0.247$\pm$0.031  &0.181$\pm$0.012  &0.201$\pm$0.016  &0.251$\pm$0.020  &0.259$\pm$0.013  &0.272$\pm$0.014  &0.272$\pm$0.019  &0.296$\pm$0.012  &0.281$\pm$0.019  &0.271$\pm$0.011 \\
&LINE&0.132$\pm$0.003  &0.153$\pm$0.010  &0.167$\pm$0.004  &0.157$\pm$0.007  &0.179$\pm$0.008  &0.186$\pm$0.007  &0.203$\pm$0.007  &0.200$\pm$0.011  &0.190$\pm$0.010  &0.221$\pm$0.013 \\
&DeepWalk&0.113$\pm$0.001  &0.117$\pm$0.002  &0.120$\pm$0.002  &0.121$\pm$0.003  &0.123$\pm$0.002  &0.121$\pm$0.001  &0.123$\pm$0.002  &0.124$\pm$0.003  &0.124$\pm$0.002  &0.123$\pm$0.002 \\
\cmidrule{1-12}
\multirow{5}{*}{\rotatebox{90}{Coverage}}
&{\our} &0.187$\pm$0.005  &\textbf{0.228}$\pm$\textbf{0.016}  &\textbf{0.264}$\pm$\textbf{0.011}  &\textbf{0.269}$\pm$\textbf{0.015}  &\textbf{0.306}$\pm$\textbf{0.011}  &\textbf{0.276}$\pm$\textbf{0.014}  &\textbf{0.318}$\pm$\textbf{0.011}  &0.258$\pm$0.002  &\textbf{0.285}$\pm$\textbf{0.013}  &\textbf{0.292}$\pm$\textbf{0.009}  \\
&{\oursingle} &\textbf{0.200}$\pm$\textbf{0.021}  &0.190$\pm$0.005  &0.224$\pm$0.008  &0.227$\pm$0.006  &0.237$\pm$0.011  &0.220$\pm$0.011  &0.229$\pm$0.004  &\textbf{0.270}$\pm$\textbf{0.006}  &0.266$\pm$0.005  &0.271$\pm$0.009\\
\cmidrule{2-12}
&Auto-encoder&0.198$\pm$0.020  &0.153$\pm$0.009  &0.167$\pm$0.011  &0.201$\pm$0.013  &0.206$\pm$0.008  &0.214$\pm$0.009  &0.213$\pm$0.012  &0.228$\pm$0.007  &0.219$\pm$0.012  &0.213$\pm$0.007 \\
&LINE&0.117$\pm$0.003  &0.133$\pm$0.008  &0.143$\pm$0.003  &0.136$\pm$0.005  &0.152$\pm$0.006  &0.157$\pm$0.005  &0.168$\pm$0.005  &0.167$\pm$0.008  &0.160$\pm$0.007  &0.181$\pm$0.009   \\
&DeepWalk&0.102$\pm$0.001  &0.105$\pm$0.001  &0.107$\pm$0.002  &0.108$\pm$0.002  &0.110$\pm$0.001  &0.108$\pm$0.001  &0.110$\pm$0.001  &0.110$\pm$0.002  &0.111$\pm$0.002  &0.110$\pm$0.001 \\
\cmidrule{1-12}
\multirow{5}{*}{\rotatebox{90}{Expansion}}
&{\our}&0.813$\pm$0.005  &\textbf{0.772}$\pm$\textbf{0.016}  &\textbf{0.736}$\pm$\textbf{0.011}  &\textbf{0.731}$\pm$\textbf{0.015}  &\textbf{0.694}$\pm$\textbf{0.011}  &\textbf{0.724}$\pm$\textbf{0.014}  &\textbf{0.682}$\pm$\textbf{0.011}  &0.742$\pm$0.002 &\textbf{0.715}$\pm$\textbf{0.013}  &\textbf{0.708}$\pm$\textbf{0.009}  \\
&{\oursingle} &\textbf{0.800}$\pm$\textbf{0.021}  &0.810$\pm$0.005  &0.776$\pm$0.008  &0.773$\pm$0.006  &0.763$\pm$0.011  &0.780$\pm$0.011  &0.771$\pm$0.004  &\textbf{0.730}$\pm$\textbf{0.006}  &0.734$\pm$0.005  &0.729$\pm$0.009 \\
\cmidrule{2-12}
&Auto-encoder&0.802$\pm$0.020  &0.847$\pm$0.009  &0.833$\pm$0.011  &0.799$\pm$0.013  &0.794$\pm$0.008  &0.786$\pm$0.009  &0.787$\pm$0.012  &0.772$\pm$0.007  &0.781$\pm$0.012  &0.787$\pm$0.007   \\
&LINE&0.883$\pm$0.003  &0.867$\pm$0.008  &0.857$\pm$0.003  &0.864$\pm$0.005  &0.848$\pm$0.006  &0.843$\pm$0.005  &0.832$\pm$0.005  &0.833$\pm$0.008  &0.840$\pm$0.007  &0.819$\pm$0.009   \\
&DeepWalk&0.898$\pm$0.001  &0.895$\pm$0.001  &0.893$\pm$0.002  &0.892$\pm$0.002  &0.890$\pm$0.001  &0.892$\pm$0.001  &0.890$\pm$0.001  &0.890$\pm$0.002  &0.889$\pm$0.002  &0.890$\pm$0.001 \\
\bottomrule

\end{tabular}\vspace{-15pt}
}
\end{table*}

The network embedding models compared in the experiments are listed as follows
\begin{itemize}

\item \textit{{\our}}: {\our} is the synergistic embedding model for multiple aligned heterogeneous networks introduced in this paper. {\our} preserves both the local and global network structure with a set of \textit{meta proximity} calculated from each of the heterogeneous network. {\our} transfers the information from the aligned mature networks to the emerging network with the anchor links, which accommodate the learnt embedding feature vectors for the anchor users in the aligned networks.

\item \textit{{\oursingle}}: {\oursingle} is a variant model of {\our} proposed in this paper, which preserves both the local and global network structure with a set of \textit{meta proximity} based on the heterogeneous networks. {\oursingle} effectively fuses the heterogeneous information inside the network, where the fusion weight of information in different categories can be learnt automatically.

\item \textit{Auto-encoder Model}: The {\autoencoder} model proposed in \cite{BLPL06} can project the instances into a low-dimensional feature space. In the experiments, we build the {\autoencoder} model merely based on the friendship link among users, where the feature vector for each user is his/her social adjacency vector. Here, we also adjust the loss term for {\autoencoder} by weighting the non-zero features more with parameter $\gamma$ as introduced in Section~\ref{subsec:single}.

\item \textit{LINE Model}: The {\linemodel} model is a scalable network embedding model proposed in \cite{TQWZYM15}, which optimizes an objective function that preserves both the local and global network structures. {\linemodel} also uses a edge-sampling algorithm to addresses the limitation of the classical stochastic gradient descent, which improves the inference effectiveness and the efficiency greatly.

\item \textit{DeepWalk Model}: The {\deepwalk} model \cite{PAS14} extends the word2vec model \cite{MSCCD13} to the network embedding scenario. {\deepwalk} uses local information obtained from truncated random walks to learn latent representations.

\end{itemize}

\subsubsection{Link Prediction Experimental Setting}

Given the emerging network, from which we can obtain all the existing links inside the network as the positive set. Meanwhile, from the network, a subset of the non-existing links are randomly sampled as the negative set according to the \textit{negative positive sampling ratio} $\theta \in \{1, 2, \cdots, 10\}$. Here $\theta = 1$ denotes that the negative set is of the same size as the positive set. Meanwhile, $\theta = 10$ represents the negative set is 9 times larger than the positive set. The positive and negative sets are divided into two subsets with 10-fold cross validation, where $9$ folds are used as the training set and $1$ fold is used as the testing set.

To denote different degrees of information sparsity, the emerging network is further sampled to remove information randomly from the network, which is controlled by the \textit{sampling ratio} $\lambda \in \{10\%, 20\%, \cdots, 100\%\}$. Here, the sampling denotes removing the positive links in the training set, as well as the posts from the emerging network to make the network sparse. $\lambda = 10\%$ denotes $10\%$ of the information is preserved ($90\%$ of the information is randomly removed); while $\lambda = 100\%$ denotes that all the information is preserved in the emerging network. The network embedding is learnt based on the training set and network after sampling. 

With the learnt embedding feature vectors, the remaining links in the training set is used to build a supervised link prediction model. For each link $(u, v)$ in the training set, the embedding feature vector of the nodes $u$ and $v$ are concatenated as the link feature vector. Depending on whether link $(u, v)$ appears in the positive set or negative set, $(u, v)$ will be assigned with the $+1$ or $-1$ label. SVM is used as the base classifier for all the embedding models. We train SVM with the training set, and then apply the trained SVM to the testing set to infer the labels and the formation probabilities of these links. By comparing the prediction labels (and inference probabilities) with the ground truth labels, we can evaluate the performance of the embedding models with different kinds of evaluation metrics to be introduced in the next subsection. 

In the experiments, $7$ hidden layers are involved in framework {\our} (3 hidden layers in encoder step, 3 in decoder step, and 1 fusion hidden layer). The number neuron in these hidden layers are $500$, $50$, $50 \times 7$, $50$, $50 \times 7$, $50$ and $500$ respectively. Epoch is $600$ and the batch size is $64$. The parameters $\alpha = 1.0$, $\beta = 0.02$ and $\gamma = 100.0$ are used in the experiments.

\subsubsection{Link Prediction Evaluation Metrics}

By comparing the link prediction results in the testing set, i.e., the inference probabilities, with the ground truth labels, the performance of different embedding models can be evaluated by AUC as the metric. Meanwhile, based on the prediction labels, we can evaluate the performance of these embedding models with Recall, F1 and Accuracy as the metrics. 

\begin{figure*}[t]
\vspace{-30pt}
\centering
\subfigure[AUC]{ \label{fig:link_prediction_parameter_analysis_1}
    \begin{minipage}[l]{0.45\columnwidth}
      \centering
      \includegraphics[width=1.0\textwidth]{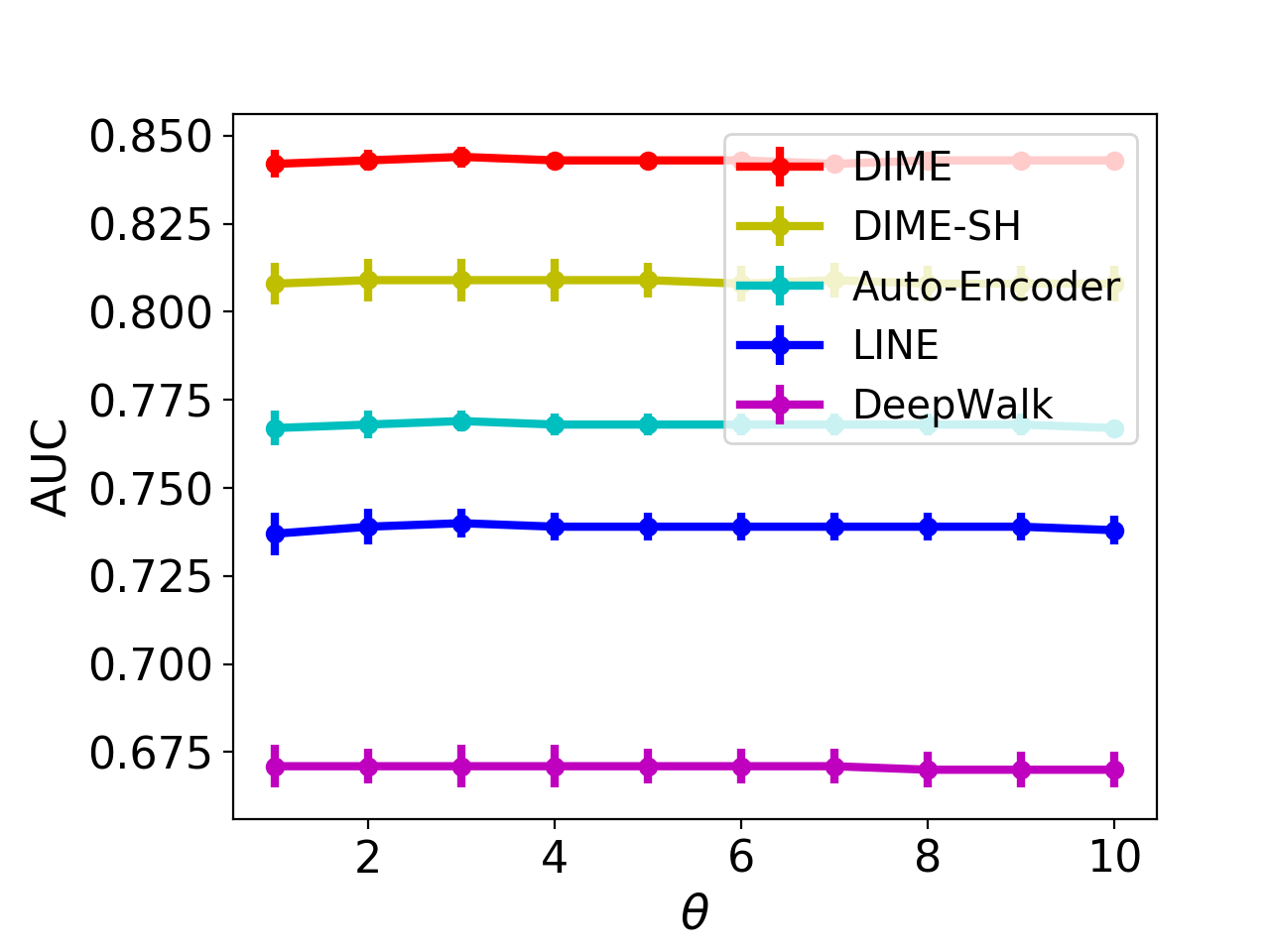}
    \end{minipage}
}
\subfigure[Recall]{\label{fig:link_prediction_parameter_analysis_3}
    \begin{minipage}[l]{0.45\columnwidth}
      \centering
      \includegraphics[width=1.0\textwidth]{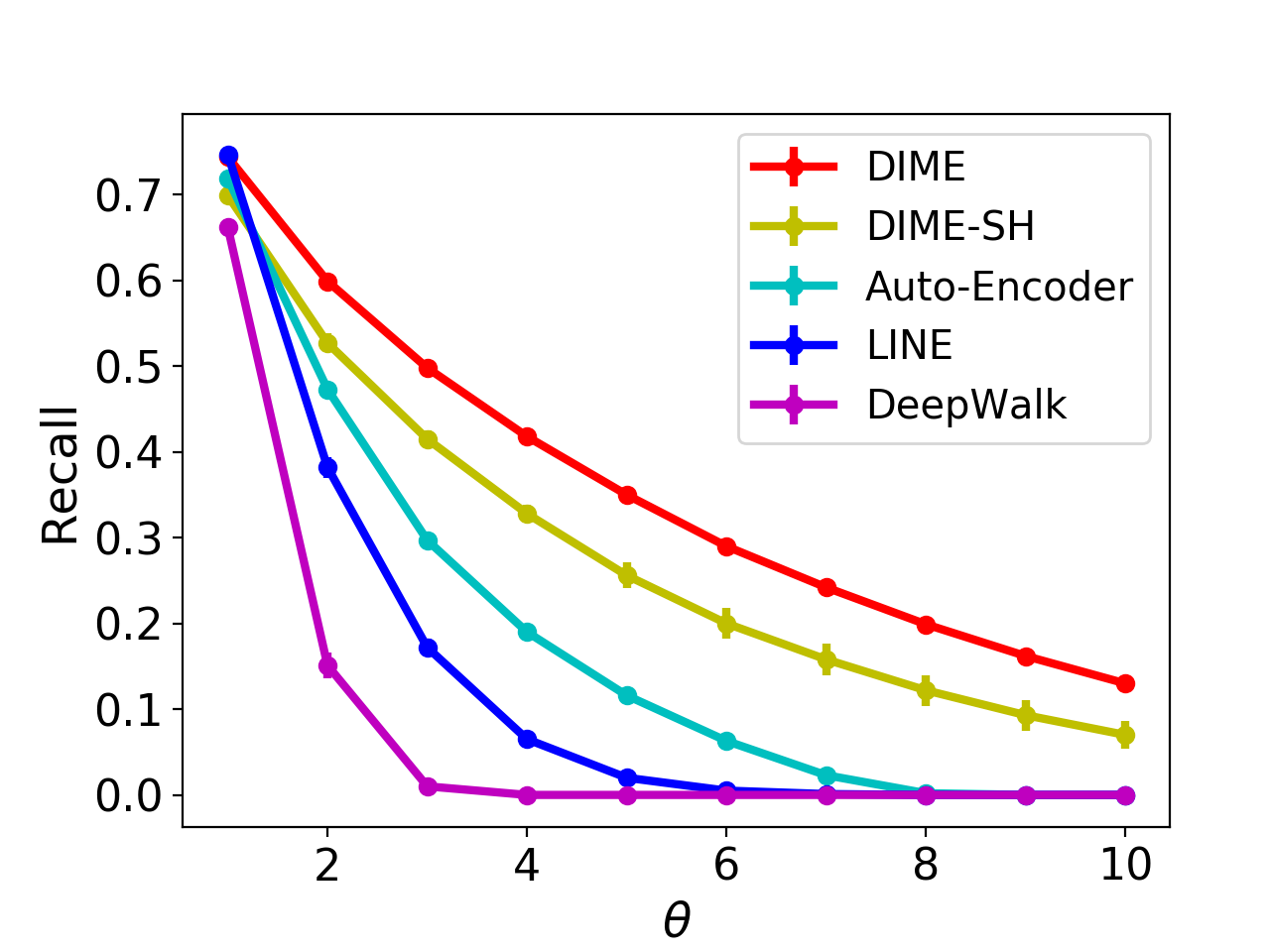}
    \end{minipage}
}
\subfigure[F1]{ \label{fig:link_prediction_parameter_analysis_4}
    \begin{minipage}[l]{0.45\columnwidth}
      \centering
      \vspace{-7pt}
      \includegraphics[width=1.0\textwidth]{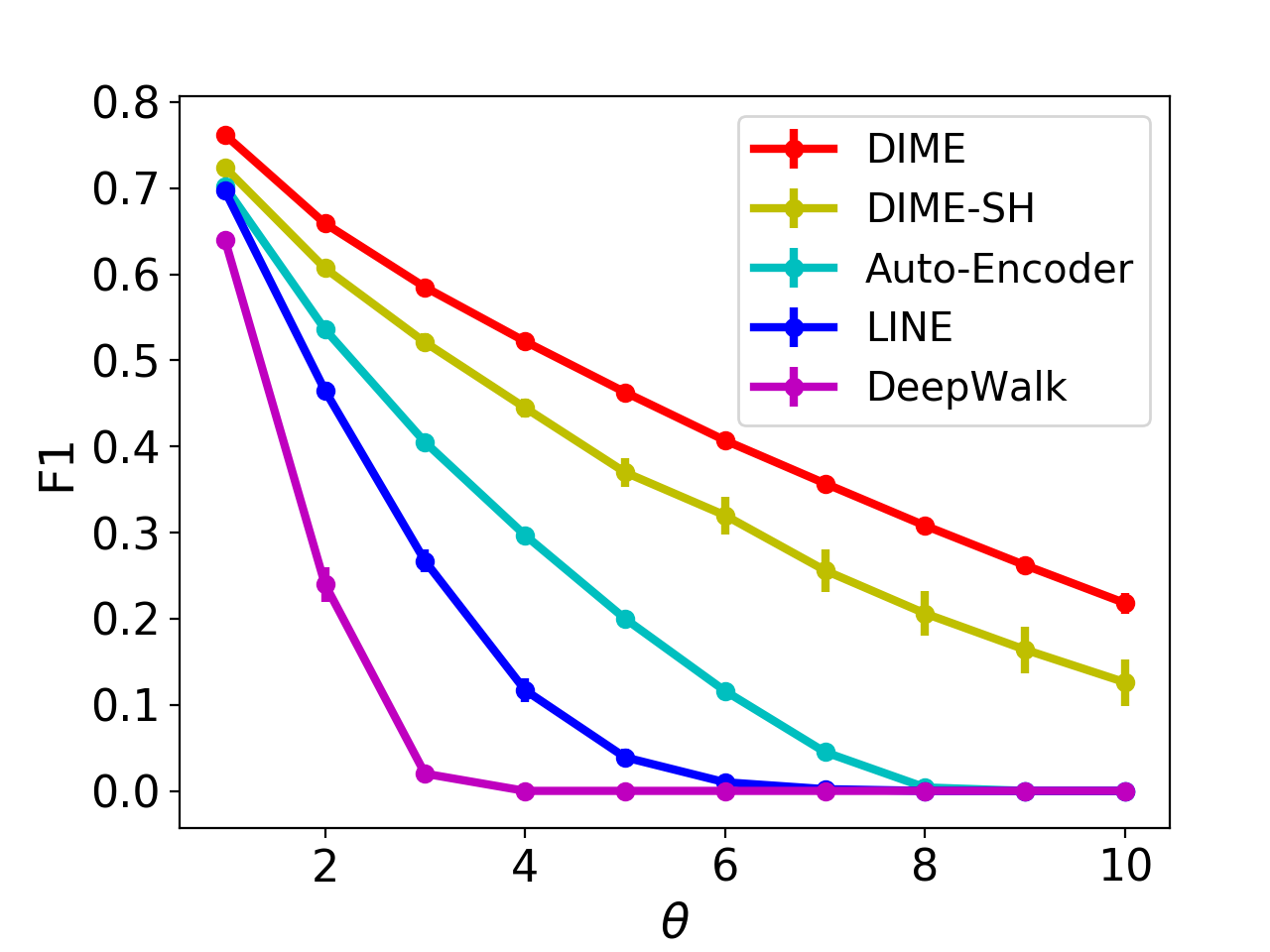}
    \end{minipage}
}
\subfigure[Accuracy]{ \label{fig:link_prediction_parameter_analysis_5}
    \begin{minipage}[l]{0.45\columnwidth}
      \centering
      \vspace{-7pt}
      \includegraphics[width=1.0\textwidth]{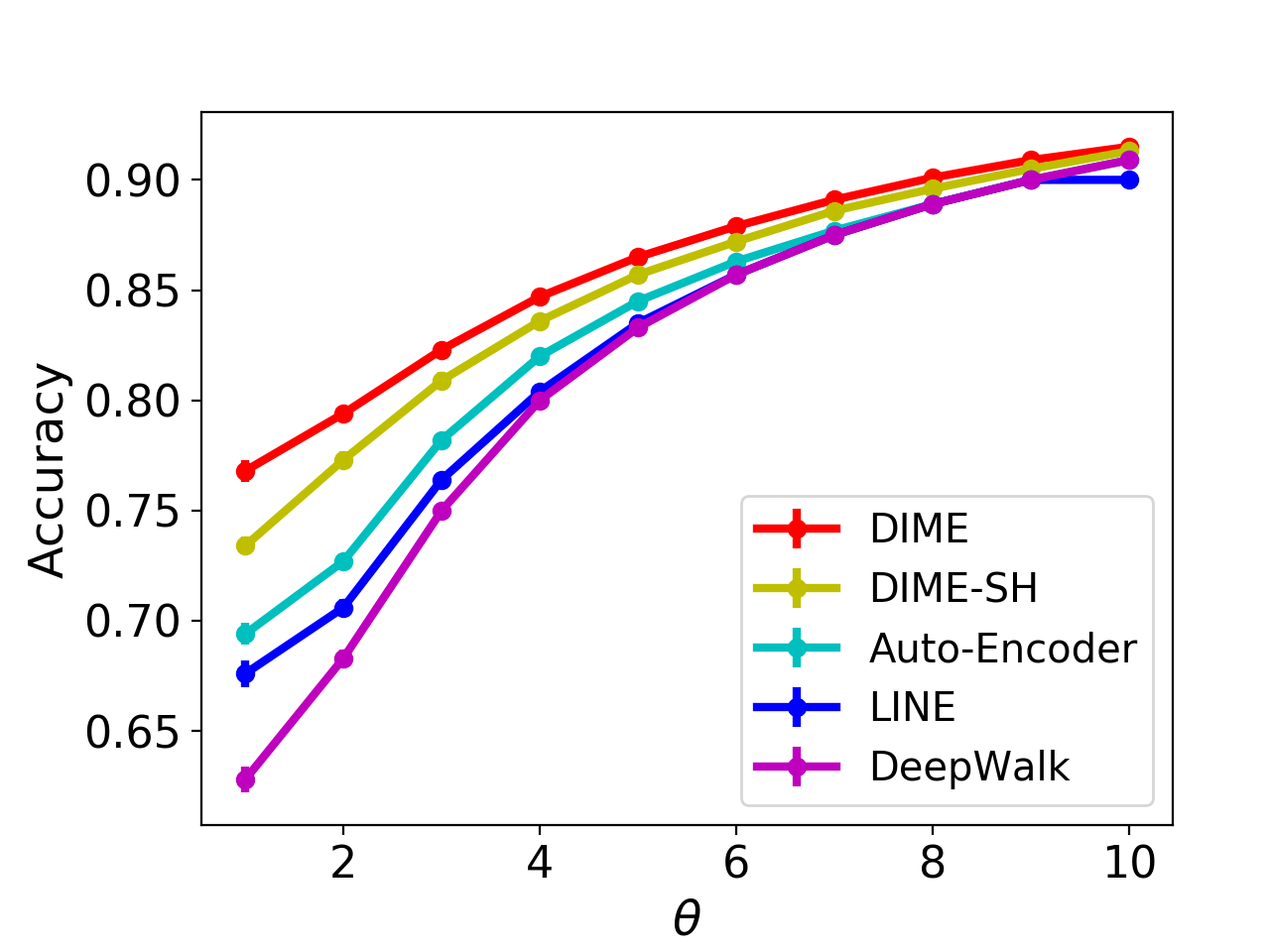}
    \end{minipage}
}\vspace{-7pt}
\caption{Parameter Analysis of negative positive rate $\theta$ in link prediction.}\label{fig:link_prediction_parameter_analysis}\vspace{-15pt}
\end{figure*}

\subsubsection{Community Detection Experimental Setting}

Different from \textit{link prediction}, \textit{community detection} is an unsupervised learning task, where no training set is needed. Based on the whole network, we randomly sample a subset of information, i.e., follow links and posts, from the network controlled by the \textit{sampling ratio} $\lambda \in \{10\%, 20\%, \cdots, 100\%\}$. Based on the sampled network, we learn the embedding of the emerging network and get the embedded feature vector for each user in the emerging network. KMeans is applied as the base clustering model to partition the users into different clusters based on their learnt embedding feature vectors. We evaluate the performance of the embedding methods by comparing the clustering results with the original network structure (involving users and follow links) before sampling. The evaluation metrics will be introduced in the following subsection.

\subsubsection{Community Detection Evaluation Metrics}

The \textit{community detection} evaluation metrics used in the experiments include, Density, Separability, Coverage, and Expansion, which have been frequently used as the metrics for topological clustering problems. An introduction to these metrics is available in \cite{AGMZ11} and \cite{JJ12}.

\subsection{Link Prediction Experimental Results}
 
 In this link prediction task, we compare the performance of five different embedding methods under different \textit{sampling ratio} $\lambda \in \{10\%, 20\%, \cdots, 100\%\}$. We try to predict the follow link relationship in the testing set with the sampled training data. The negative positive rate $\theta$ is set with value $1$ here (i.e., negative and positive sets are of the same size).
 
The method we proposed in this paper, {\our}, performs much better than the other methods in the link prediction task, since the heterogeneous information from both the emerging and other aligned mature networks used in {\our} can provide extra information to help the model learn the embedding feature vectors of the users. Table~\ref{tab:link_prediction_result} shows the performance of {\our}, {\oursingle}, and other three baseline methods, Auto-eocoder, LINE and DeepWalk, evaluated by AUC, Accuracy, Recall and F1 with different sampling ratio $\lambda$s. 

When the \textit{sampling ratio} $\lambda$ is low, like $10\%$, the baseline models will suffer from the information sparsity a lot, but by transferring information other aligned source networks {\our} can still obtain very good performance. As the \textit{sampling ratio} $\lambda$ increases, the performance of all these methods improves steadily, and {\our} can outperform the other methods with great advantages consistently.   

Among all the baseline methods, {\our} can achieve the best performance in most of the cases (except the Recall measure with $\lambda \in \{10\%, 20\%, 30\%, 40\%\}$). For instance, when $\lambda = 30\%$, the AUC achieved by {\our} is $0.838$, which is $4.5\%$ higher than the AUC obtained by {\oursingle}. It demonstrates our assumption that ``information from other aligned networks can help improve the performance greatly''. The advantages of {\our} will be more significant compared with the remaining baseline methods. Meanwhile, with heterogeneous information in the emerging network, {\oursingle} can also outperform the other baseline models built with homogeneous information only. For instance, the AUC, Accuracy, and F1 obtained by {\oursingle} are all over $8\%$ greater than the measures obtained by Auto-encoder, LINE and DeepWalk. It shows the meta proximity proposed in this paper can effectively capture the network structure information for the users.

\begin{figure*}[t]
\vspace{-30pt}
\centering
\subfigure[Density]{ \label{fig:community_detection_density}
    \begin{minipage}[l]{0.45\columnwidth}
      \centering
      \includegraphics[width=1.0\textwidth]{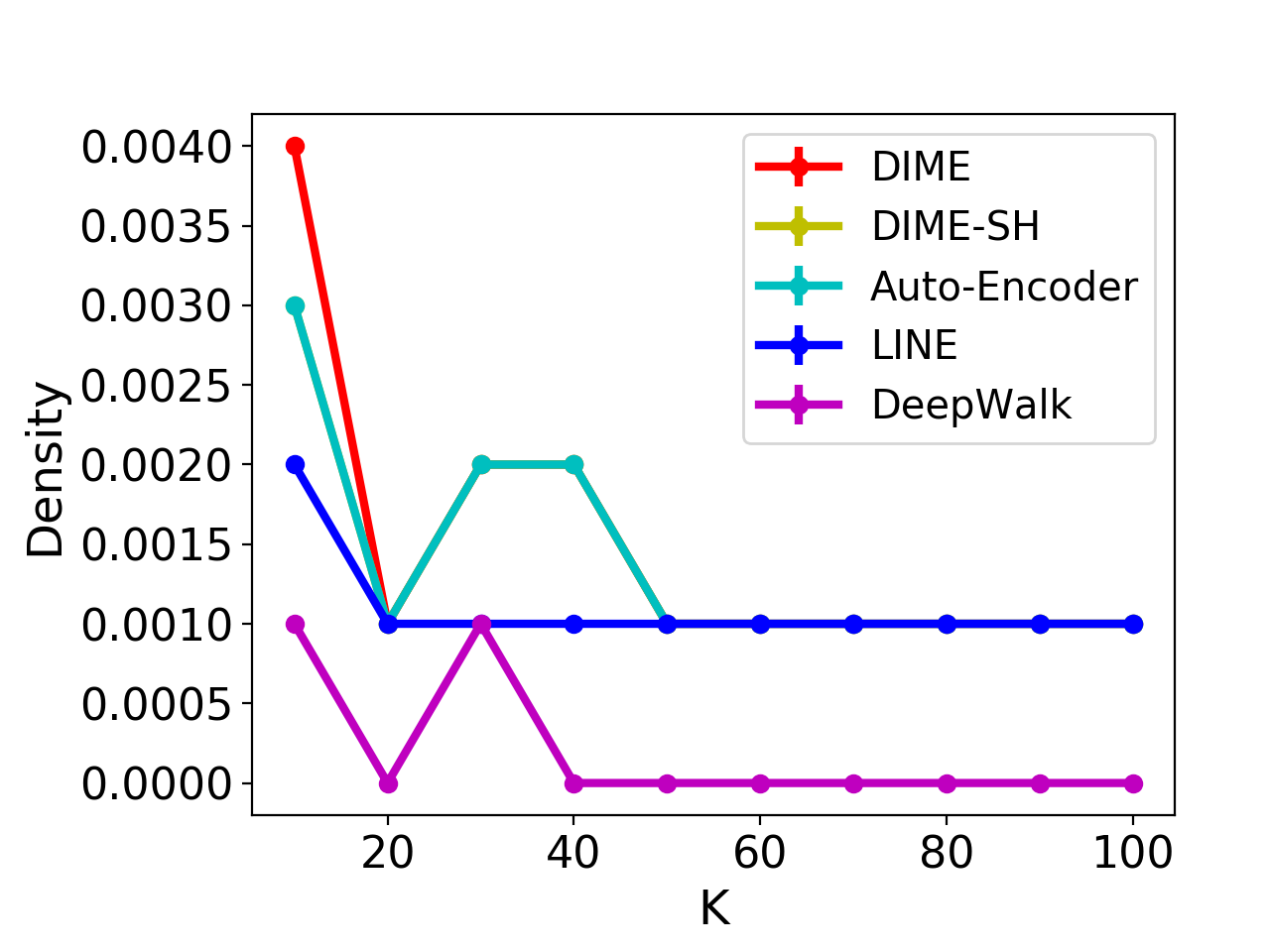}
    \end{minipage}
}
\subfigure[Separability]{ \label{fig:community_detection_separability}
    \begin{minipage}[l]{0.45\columnwidth}
      \centering
      \includegraphics[width=1.0\textwidth]{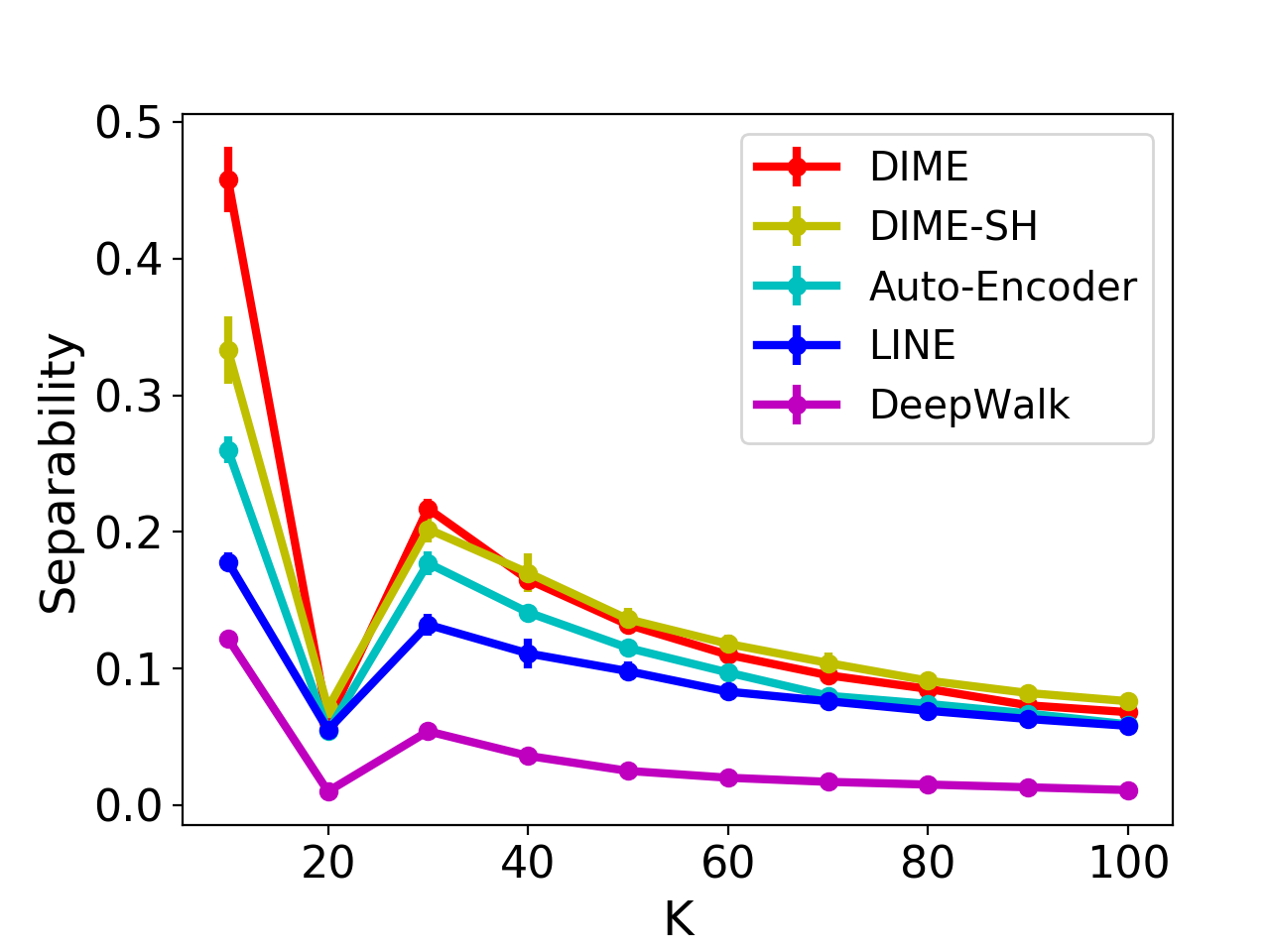}
    \end{minipage}
}
\subfigure[Coverage]{\label{fig:community_detection_separability}
    \begin{minipage}[l]{0.45\columnwidth}
      \centering
      \vspace{-7pt}
      \includegraphics[width=1.0\textwidth]{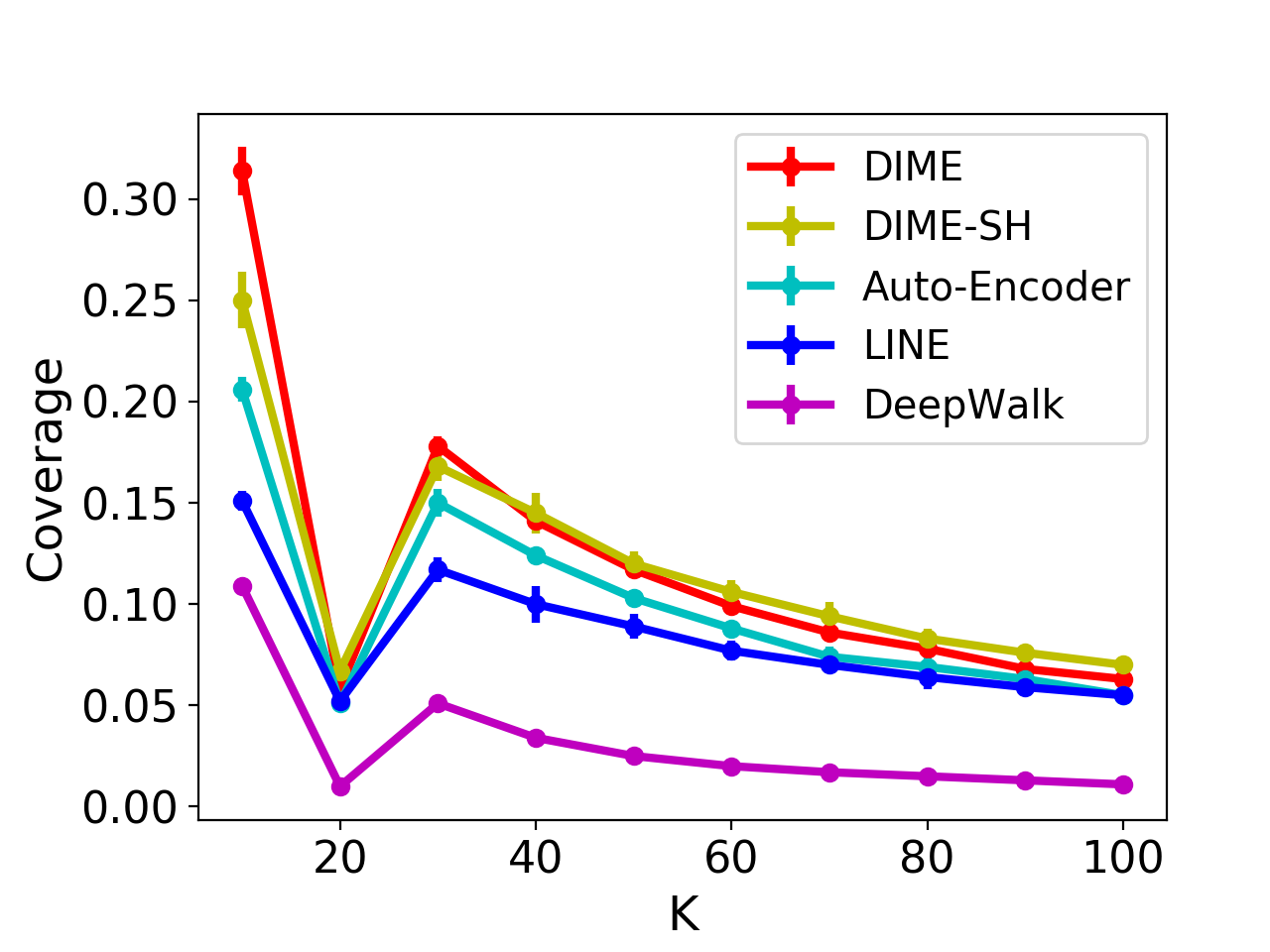}
    \end{minipage}
}
\subfigure[Expansion]{ \label{fig:community_detection_expansion}
    \begin{minipage}[l]{0.45\columnwidth}
      \centering
      \vspace{-7pt}
      \includegraphics[width=1.0\textwidth]{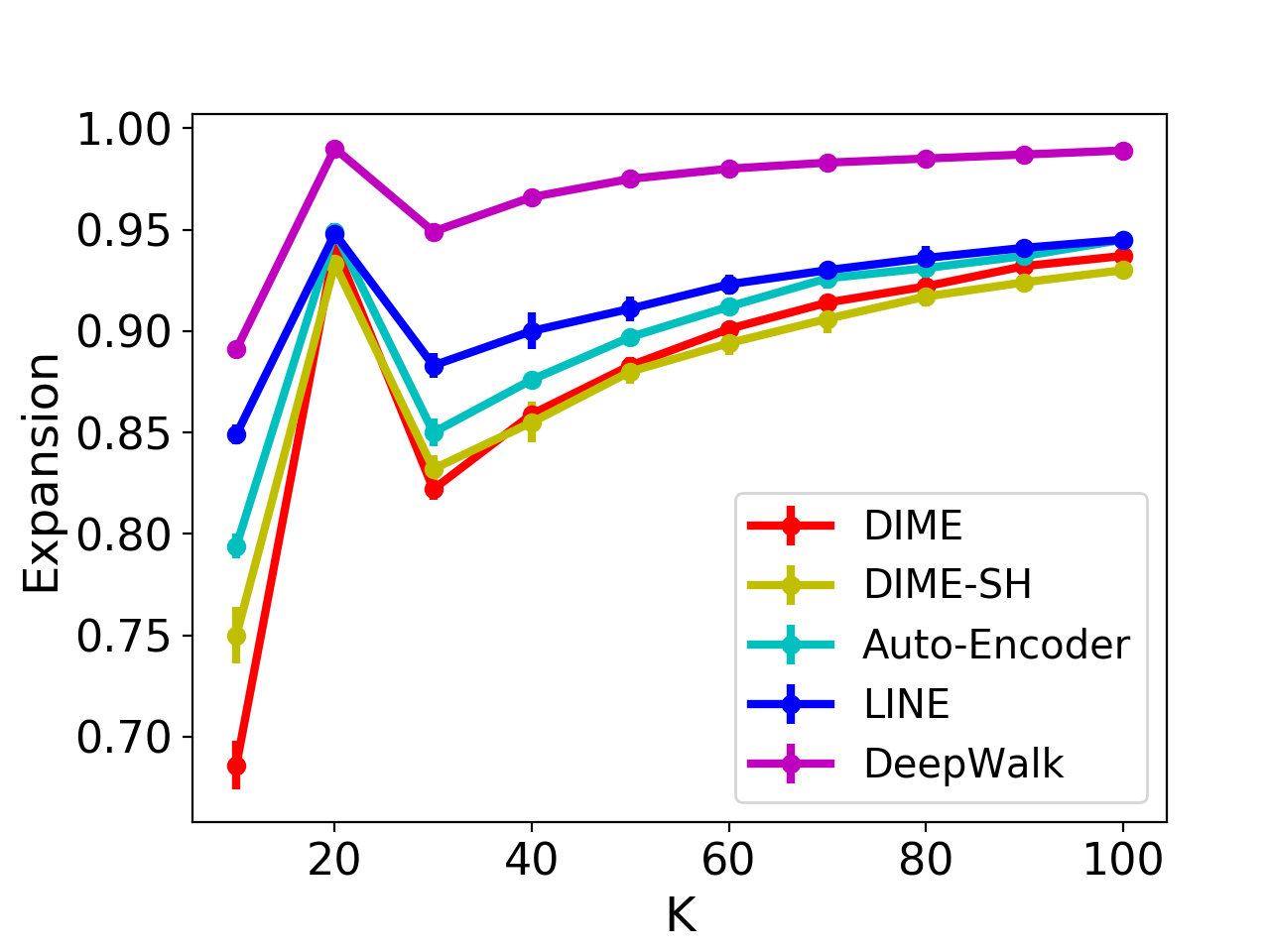}
    \end{minipage}
}
\vspace{-7pt}
\caption{Parameter Analysis of community number k.}\label{fig:community_detection_parameter_analysis}\vspace{-15pt}
\end{figure*}


\subsection{Link Prediction Parameter Sensitivity Analysis}

In the link prediction task, we set the negative positive rate $\theta$ equals 1. In this part, we will provide the sensitivity analysis of parameter $\theta$. Figure~\ref{fig:link_prediction_parameter_analysis} shows the AUC, Recall, F1 and Accuracy of the comparison methods with negative positive sample rate $\theta \in \{1, 2, \cdots, 10\}$.

As the negative positive rate $\theta$ increases, more negative links will be added to the training and testing set, which renders the link prediction task more challenging. According to Figure~\ref{fig:link_prediction_parameter_analysis}, we observe that, the metrics like Recall and F1 are all decreasing as $\theta$ gets larger. The AUC curve is relative stable, as it is not very sensitive to the class imbalance problem. As for Accuracy, when the negative positive rate $\theta$ increases, the data gets more imbalanced. In such a class imbalance circumstance, the high Accuracy scores will not be that meaningful.

Through Figure~\ref{fig:link_prediction_parameter_analysis}, we can observe that even the metrics like Recall and F1 will all degrade as the the negative positive rate $\theta$ increase, the decreasing speed of different methods is different. {\our} decreases slower than {\oursingle}, while the decreasing speed of {\oursingle} is much slower than the remaining baseline methods. This means that although the performance of all the methods are influenced by the increasing parameter $\theta$, {\our} and {\oursingle} are more stable than the other baseline models.

\subsection{Community Detection Experimental Results}
In Table~\ref{tab:community_detection_result}, we show the community detection results obtained by the comparison methods evaluated by Density, Separability, Coverage and Expansion with different sampling ratios $\lambda \in \{10\%, 20\%, \cdots, 100\%\}$. Here the number of cluster, i.e., parameter $k$, is assigned with value $10$, whose sensitivity analysis will be provided in the following subsection.

According to the results shown in the table, we observe that {\our} performs the best out of all the methods. As we can see, when the sample ratio increases, the emerging network will have more information, and the community detection results obtained by all the comparison methods will increase steadily.

Under the same sample ratio, the comparison methods sorted in the descending order according to their performance are as follows: {\our}, {\oursingle}, Auto-encoder, LINE and DeepWalk. By comparing {\our} with the other baseline methods, {\our} can outperform them with great advantages. For instance, when sample rate $\lambda$ equals 0.5, the Separability of {\our} is 0.440, which is $41.5\%$ larger than that obtained by {\oursingle}. Based on the meta proximity and heterogeneous information in the emerging network, {\oursingle} can perform much better than the other homogeneous network based embedding methods. For instance, the Separability achieved by {\oursingle} is $20\%$ larger than that of Auto-encoder, and over $70\%$ greater than LINE and DeepWalk. The results are also very similar for other evaluation metrics.

\subsection{Community Detection Parameter Sensitivity Analysis}

In the community detection task, we set the community number $k$ with 10. In this part, we try to analyze how will the performance be influenced while the number of communities $k$ differs. Figure~\ref{fig:community_detection_parameter_analysis} shows the change of Density, Separability, Coverage, Expansion obtained by the comparison methods while $k$ increases from 10 to 100.

Generally, when the community number $k$ increases from 10 to 20, the performance of all the methods degrades a little bit, and when k increases from 20 to 30, the performance increases again, which will keep dropping steadily as $k$ further increases. In the community detection task, we do not know how many communities in there. So we need to try different $k$s to get best performance. In our case, $k=10$ achieves best performance among the values in $\{10, 20, \cdots, 100\}$.

If we sort the comparison methods according to their performance in the decreasing order, the sorted list will be {\our}, {\oursingle}, Auto-encoder, LINE and DeepWalk. The heterogeneous information across the emerging and aligned source networks used in {\our} help the clustering model to group similar people together.

\section{Related Work} \label{sec:related_work}


Network embedding has become a very hot research problem in recent years, which can project a graph-structured data to the feature vector representations automatically. In the graphs, the relation can be treated as a translation of the entities, and many translation based embedding models have been proposed. Model TransE \cite{BUGWY13} is the initial translation based embedding work, which projects the entity and relation into a common feature space. TransH \cite{WZFC14} improves TransE by considering the link cardinality constraint in the embedding process, and can achieve comparable time complexity. In the real-world multi-relational networks, the entities can have multiple aspects, and the different relations can express different aspects of the entity. Model TransR \cite{LLSLZ15} proposes to build the entity and relation embeddings in separate entity and relation spaces instead.

In recent years, many recent network embedding works based on random walk model and deep learning models have proposed, like Deepwalk \cite{PAS14}, LINE \cite{TQWZYM15}, node2vec \cite{GL16}, HNE \cite{CHTQAH15}. Perozzi et al. extends the word2vec \cite{MSCCD13} to the network scenario and introduce the Deepwalk algorithm \cite{PAS14}, which uses local information obtained from truncated random walks to learn latent representations by treating walks as the equivalent of sentences. Tang et al. \cite{TQWZYM15} propose to embed the networks with LINE algorithm, which can preserve both the local and global network structures. An edge-sampling algorithm is applied in LINE that addresses the limitation of the classical stochastic gradient descent and improves both the effectiveness and the efficiency of the inference. Grover et al. \cite{GL16} introduces a flexible notion of a node's network neighborhood and design a biased random walk procedure to sample the neighbors in the training process, which efficiently explores diverse neighborhoods. Chang et al. \cite{CHTQAH15} learns the embedding of heterogeneous networks involving both text and image information. Chen et al. \cite{CS16} introduce a task guided embedding model to learning the representations for the author identification problem.

Link prediction and recommendation first proposed in \cite{LK03} has become a very important problem in online social networks, which provides social network researchers with the opportunity to study both the network properties from the individuals social connection perspective. Traditional unsupervised link predictor proposed in \cite{LK03} mainly calculate the closeness scores among users, and assume that close users tend to be friends in the network. Hasan et al. \cite{HCSZ06} is the first to study the link prediction problem as a supervised learning problem, where the existing and non-existing social links are treated as the positive and negative instances respectively. Today, many social networks are heterogeneous and to conduct the link prediction in these networks, Sun et al. \cite{SBGAH11} propose a meta path-based prediction model to predict co-author relationship in the heterogeneous bibliographic network.


Clustering method has also been widely used to detect communities in networks. Newman et al. introduce a modularity function measuring the quality of a division of networks \cite{NG04}. Shi et al. introduce the concept of normalized cut and discover that the eigenvectors of the Laplace matrix provide a solution to the normalized cut objective function \cite{SM00}. In addition, many community detection works have been done on heterogeneous online social networks. Sun et al. \cite{SAH12} propose to study the clustering problem with complete link information but incomplete attribute information. Lin et al. \cite{LKYWJL12} try to detect the communities in networks with incomplete relational information but complete attribute information. 

\section{Conclusion}\label{sec:conclusion}

In this paper, we propose to study the embedding problem for emerging online social networks with broad learning, namely the {\problem} problem. Emerging networks denote the social networks that newly created containing very little social information. Traditional embedding models will suffer from the information sparsity problem a lot in handling such emerging networks. To solve problem, we introduce a novel embedding framework {\our}. Based on a set of meta proximity, {\our} can make full use of the heterogeneous information inside the network. Via the cross-network information transfer, {\our} refines the embedding results with information from other external aligned mature networks. To demonstrate the effectiveness of {\our}, extensive experiments have been done on real-world social networks, which include two main tasks: link prediction and community detection. The experimental results show that {\our} can perform very well in learning the embedding vectors for nodes in the emerging networks.

\balance
\bibliographystyle{plain}
\bibliography{reference}

\end{document}